# Bayesian System Identification based on Hierarchical Sparse Bayesian Learning and Gibbs Sampling with Application to Structural Damage Assessment

Yong Huang[a,b], James L. Beck[b,*] and Hui Li[a]

[a]Key Lab of Structural Dynamic Behavior and Control of the Ministry of Education, School of Civil Engineering, Harbin Institute of Technology, Harbin, China

[b]Division of Engineering and Applied Science, California Institute of Technology, CA, USA

**ABSTRACT:**

Bayesian system identification has attracted substantial interest in recent years for inferring structural models based on measured dynamic response from a structural dynamical system. The focus in this paper is Bayesian system identification based on noisy incomplete modal data where we can impose spatially-sparse stiffness changes when updating a structural model. To this end, based on a similar hierarchical sparse Bayesian learning model from our previous work, we propose two Gibbs sampling algorithms. The algorithms differ in their strategies to deal with the posterior uncertainty of the equation-error precision parameter, but both sample from the conditional posterior probability density functions (PDFs) for the structural stiffness parameters and system modal parameters. The effective dimension for the Gibbs sampling is low because iterative sampling is done from only three conditional posterior PDFs that correspond to three parameter groups, along with sampling of the equation-error precision parameter from another conditional posterior PDF in one of the algorithms where it is not integrated out as a "nuisance" parameter. A nice feature from a computational perspective is that it is not necessary to solve a nonlinear eigenvalue problem of a structural model. The effectiveness and robustness of the proposed algorithms are illustrated by applying them to the IASE-ASCE Phase II simulated and experimental benchmark studies. The goal is to use incomplete modal data identified before and after possible damage to detect and assess spatially-sparse stiffness reductions induced by any damage. Our past and current focus on meeting challenges arising from Bayesian inference of structural stiffness serve to strengthen the capability of vibration-based structural system identification but our methods also have much broader applicability for inverse problems in science and technology where system matrices are to be inferred from noisy partial information about their eigenquantities.

## 1. INTRODUCTION

Inverse problems are a core part of structural system identification, which is concerned with the determination of structural models and their parameters (e.g., stiffness) based on measured structural dynamic response (e.g., [1-5]).

*Corresponding author.
*E-mail address:* jimbeck@caltech.edu (James L. Beck)

However, real inverse problems when treated deterministically are typically ill-conditioned and often ill-posed when using noisy incomplete data, i.e., uniqueness, existence and robustness to noise of an inverse solution is not guaranteed. To deal with ill-conditioning and ill-posedness in inverse problems, a common approach is to avoid an explicit treatment of uncertainty and employ a least-squares approach with the addition of a regularization term to the data-matching term in the objective function to be optimized, often called Tikhonov Regularization (e.g., [6-7]). A regularized least-squares approach usually leads to a well-conditioned and well-posed deterministic optimization problem; however, the relationship of the unique solution to an exact solution of the original inverse problem is uncertain.

The values of parameters of structural models used to predict structural behavior are uncertain. Indeed, since these models are based on simplifying and approximate assumptions, there are really no true values of the model parameters. These modelling uncertainties suggest that when solving inverse problems, we should not just search for a single "optimal" parameter vector to specify the structural model, but rather attempt to describe the family of all plausible values of the model parameter vector that are consistent with both the observations and our prior information. This leads us to consider inverse problems in structural system identification from a full Bayesian perspective, which provides a robust and rigorous framework due to its ability to account for model uncertainties ([1, 8-11]). The posterior probability distribution of the model parameters from Bayes Theorem is used to quantify the plausibility of all models, within a specified set of models, based on the available data. As an aside, we note that any Tikhonov-regularized solution can always be viewed as the MAP (maximum a posteriori) solution of a Bayesian inversion where the log likelihood function and log prior distribution come from the data-fit and regularization terms, respectively. Oh [12] lists broad categories of prior PDFs for Bayesian inverse problems and compared their corresponding model regularization effects.

A general strategy to alleviate ill-conditioning and ill-posedness in inverse problems is to incorporate as much prior knowledge as possible in a Bayesian formulation in order to constrain the set of plausible solutions. For example, in structural health monitoring, the engineer may wish to exploit the prior knowledge that structural stiffness change from damage typically occurs at a limited number of locations in a structure in the absence of its collapse. We have previously proposed some Bayesian methods for sparse stiffness loss inversions [13-16]. These methods build on developments in Sparse Bayesian Learning [17] and its application in Bayesian compressive sensing [18,19], where sparseness of the inferred structural stiffness loss is promoted automatically, allowing more robust damage localization with higher-resolution. Our methods have general applicability for Bayesian system identification, however. They can be used, for example, to find sparse adjustments of the stiffness distribution specified by a finite-element structural model so that it becomes more consistent with the measured response of the structure. Alternatively, if sufficient data are available, our methods allow a non-sparse set of stiffness changes to be inferred.

Our methods are based on a specific hierarchical Bayesian model where the Bayesian analysis at the lowest level can be done analytically because it involves Gaussian distributions. However, the higher levels involve the evaluation of multi-dimensional integrals over the hyper-parameter space of the Bayesian model that are analytically intractable. Laplace's method of asymptotic approximation [20] was used in our previously proposed sparse Bayesian learning methods [13-15]. These approximations ignore the posterior uncertainty in the hyper-parameters and system modal parameters, however, so a goal of the work presented here is to provide a fuller treatment of the posterior uncertainty by employing Markov chain Monte Carlo (MCMC) simulation methods [21], so that some of the Laplace approximations can be avoided.

In recent years, MCMC methods have received much attention for Bayesian model updating in which samples consistent with the posterior PDF of the model parameters are generated. An advantage of MCMC methods is that they can provide a full characterization of the posterior uncertainty, even when the model class is not globally identifiable [22,23]. Numerous new MCMC methods have been proposed over the last decade or so with the goal of improving the computational efficiency of posterior sampling in Bayesian structural model updating (e.g., [22, 24-28]). However, most existing methods are only efficient for lower dimensional problems.

In this paper, we explore the use of Gibbs sampling (GS) [29,30], a special case of Markov chain Monte Carlo simulation, to efficiently sample the posterior PDF of the high-dimensional uncertain parameter vectors that arise in our sparse stiffness identification problem. The effective dimension is kept low by decomposing the uncertain parameters into a small number of groups and iteratively sampling the posterior distribution of one parameter group conditional on the other groups and the available data. Based on different strategies to deal with the posterior uncertainty of the equation error precision parameter, two Gibbs Sampling algorithms are presented. For each algorithm, analytical expressions for the full conditional posterior PDFs are derived in order to implement the GS methods. The effectiveness of the proposed Bayesian system identification algorithms for sparse stiffness loss inversion is illustrated with simulated and experimental data examples from the IASC-ASCE Phase II benchmark studies.

## 2. BAYESIAN SYSTEM IDENTIFICATION BASED ON MODAL DATA

### *2.1. Linear structural model class and structural stiffness scaling parameters*

We consider a structure of interest and assume that modal identification is performed with low-amplitude vibration data where the structural behavior is well approximated by linear dynamics with classical normal modes for damage detection purposes. Under this hypothesis, a damping matrix need not be explicitly modeled since it does not affect the model mode shapes.

We take a class of linear structural models that has $N_d$ DOF (degrees of freedom), a known mass matrix $\mathbf{M}$ based on structural drawings and an uncertain stiffness matrix $\mathbf{K}$ that is represented as a linear combination of $(N_\theta + 1)$ substructure stiffness matrices $\mathbf{K}_{j,}\ j = 0,1, \dots N_\theta$, as follows:

$$\mathbf{K}(\boldsymbol{\theta})=\mathbf{K}_0+\sum_{j=1}^{N_\theta}\theta_j\mathbf{K}_j \tag{1}$$

The nominal substructure stiffness matrices $\mathbf{K}_j \in \mathbb{R}^{N_d\times N_d}$ represent the contribution of the $j^{ih}$ substructure to the overall stiffness matrix $\mathbf{K}$ (e.g. from a finite-element model) and $\boldsymbol{\theta} = [\theta_1, \theta_2, \dots, \theta_{N_\theta}] \in \mathbb{R}^{N_\theta}$ are corresponding stiffness scaling parameters that are learned from the modal data. For example, the substructuring may be chosen to focus on "hot spots" where the prior probability of damage occurring is higher.

### *2.2. Modal data and system modal parameters*

We assume that the target structure is instrumented with sensors measuring structural vibrations at $N_o$ DOF. Suppose that $N_s$ sets of vibration time histories are measured and $N_m$ dominant modes of the system are identified from each set of time histories. The MAP (maximum a posteriori) estimates are taken from Bayesian modal identification (e.g., [11]) to serve as the "measured" natural frequencies and mode shapes $\hat{\boldsymbol{\omega}}^2 = [\hat{\omega}_{1,1}^2, \dots, \hat{\omega}_{1,N_m}^2, \hat{\omega}_{2,1}^2, \dots, \hat{\omega}_{N_s,N_m}^2]^T$ and $\hat{\boldsymbol{\psi}} = \left[\hat{\boldsymbol{\psi}}_{1,1}^T, \dots, \hat{\boldsymbol{\psi}}_{1,N_m}^T, \hat{\boldsymbol{\psi}}_{2,1}^T, \dots, \hat{\boldsymbol{\psi}}_{N_s,N_m}^T\right]^T$, respectively, where $\hat{\omega}_{r,i}^2$ and $\hat{\boldsymbol{\psi}}_{r,i} \in \mathbb{R}^{N_o}$ give the identified modal frequency squared and a vector of the observed mode shape components of the $i^{th}$ mode ($i = 1, \dots, N_m$) from the $r^{th}$ data time segment ($r = 1, \dots, N_s$), respectively.

In typical situations where $N_o < N_d$, so only data from a subset of the DOF corresponding to the structural model are available, it is advantageous to introduce *system mode shapes* $\boldsymbol{\phi} = \left[\boldsymbol{\phi}_1^T, \dots, \boldsymbol{\phi}_{N_m}^T\right]^T \in \mathbb{R}^{N_d N_m \times 1}$ [30-33], as well as *system natural frequencies* $\boldsymbol{\omega}^2 = \left[\omega_1^2, \dots, \omega_{N_m}^2\right]^T \in \mathbb{R}^{N_m \times 1}$, to represent the actual underlying modal parameters of the linear dynamics of the structural system at all $N_d$ DOF.

### *2.3 Hierarchical Bayesian modeling*

Hierarchical Bayesian modeling is an important concept for Bayesian inference [34], which provides the flexibility to allows all sources of uncertainty and correlation to be learned from the data, and hence potentially produce more reliable system identification results. It has been used recently in Baysian system identification [35-39] where the hierarchical nature is primarily to do with the modeling of the likelihood function. To demonstrate the idea, a graphical hierarchical model representation of the structural system identification problem is illustrated in Figure 1, where the details of the Bayesian modeling are introduced in the next subsections.

#### *2.3.1 Prior for system modal parameters and structural stiffness scaling parameters*

In our modeling, the system modal parameters $\boldsymbol{\phi}$ and $\boldsymbol{\omega}^2$ are not constrained to be exact eigenvectors and eigenvalues corresponding to any structural model because there will always be modeling errors, so for $i = 1, \dots, N_m$:

$$\left(\mathbf{K}(\boldsymbol{\theta}) - \omega_i^2 \mathbf{M}\right)\boldsymbol{\phi}_i = \mathbf{e}_i \tag{2}$$

where the prior PDF for the uncertain equation errors $\mathbf{e}_i \in \mathbb{R}^{N_d}$ is modeled as $p(\mathbf{e}_i) = \mathcal{N}\left(\mathbf{e}_i | 0, \beta^{-1}\mathbf{I}_{N_d}\right)$, where $\beta$ is the common equation-error precision parameter. This maximum entropy probability model [40] gives the largest uncertainty for the set of $N_m$ $\mathbf{e}_i's$ subject to the first two moment constraints: $\mathbf{E}[(\mathbf{e}_i)_k] = 0, \mathbf{E}[(\mathbf{e}_i)_k^2] = \beta^{-1}, k = 1, \dots, N_d$. With this choice in conjunction with Eq. (2), we construct the joint prior PDF of the system modal parameters $\boldsymbol{\phi}$ and $\boldsymbol{\omega}^2$ and the stiffness scaling parameter $\boldsymbol{\theta}$ as:

$$p(\boldsymbol{\phi}, \omega^2, \boldsymbol{\theta} | \beta) = c(2\pi\beta^{-1})^{-N_m N_d/2} \exp\left\{-\frac{\beta}{2}\sum_{i=1}^{N_m}\left\|\left(\mathbf{K}(\boldsymbol{\theta}) - \omega_i^2\mathbf{M}\right)\boldsymbol{\phi}_i\right\|^2\right\} \tag{3}$$

where $c$ is a normalizing constant. The introduction of Gaussian errors in the model eigenequations means that they provide a *soft constraint* instead of imposing a rigid constraint. When the precision parameter $\beta \rightarrow \infty$, the system modal parameters become tightly clustered around the values corresponding to the structural model specified by $\boldsymbol{\theta}$, which are given by Eq. (2) with all $\mathbf{e}_i = \mathbf{0}$.

Following [15], we can decompose the joint prior PDF $p(\boldsymbol{\phi}, \omega^2, \boldsymbol{\theta}|\beta)$ into the product of a marginal PDF and a conditional PDF as in each of the following forms:

$$p(\boldsymbol{\phi}, \omega^2, \boldsymbol{\theta}|\beta) = p(\boldsymbol{\phi}, \boldsymbol{\omega}^2|\beta)p(\boldsymbol{\theta}|\boldsymbol{\phi}, \boldsymbol{\omega}^2, \beta) \tag{4a}$$

$$p(\boldsymbol{\phi}, \omega^2, \boldsymbol{\theta}|\beta) = p(\boldsymbol{\phi}, \boldsymbol{\theta}|\beta)p(\boldsymbol{\omega}^2|\boldsymbol{\theta}, \boldsymbol{\phi}, \beta) \tag{4b}$$

$$p(\boldsymbol{\phi}, \omega^2, \boldsymbol{\theta}|\beta) = p(\boldsymbol{\omega}^2, \boldsymbol{\theta}|\beta)p(\boldsymbol{\phi}|\boldsymbol{\omega}^2, \boldsymbol{\theta}, \beta) \tag{4c}$$

where the marginal prior PDFs are expressed as:

$$p(\boldsymbol{\phi}, \boldsymbol{\omega}^2|\beta) = \int p(\boldsymbol{\phi}, \boldsymbol{\omega}^2, \boldsymbol{\theta}|\beta)d\boldsymbol{\theta} = c(2\pi/\beta)^{(N_\theta - N_d N_m)/2}|\mathbf{H}^T\mathbf{H}|^{-1/2}\exp\left\{-\frac{\beta}{2}(\mathbf{b}^T\mathbf{b} - \mathbf{b}^T\mathbf{H}\,(\mathbf{H}^T\mathbf{H})^{-1}\mathbf{H}^T\mathbf{b})\right\} \tag{5a}$$

$$p(\boldsymbol{\phi}, \boldsymbol{\theta}|\beta) = \int p(\boldsymbol{\phi}, \boldsymbol{\omega}^2, \boldsymbol{\theta}|\beta)d\boldsymbol{\omega}^2 = c(2\pi/\beta)^{(N_m - N_d N_m)/2}|\mathbf{G}^T\mathbf{G}|^{-1/2}\exp\left\{-\frac{\beta}{2}(\mathbf{c}^T\mathbf{c} - \mathbf{c}^T\mathbf{G}\,(\mathbf{G}^T\mathbf{G})^{-1}\mathbf{G}^T\mathbf{c})\right\} \tag{5b}$$

$$p(\boldsymbol{\omega}^2, \boldsymbol{\theta}|\beta) = \int p(\boldsymbol{\omega}^2, \boldsymbol{\phi}, \boldsymbol{\theta}|\beta)d\boldsymbol{\phi} = c|\mathbf{F}^T\mathbf{F}|^{-1/2} \tag{5c}$$

and then the conditional prior PDFs for $\boldsymbol{\theta}$, $\boldsymbol{\omega}^2$ or $\boldsymbol{\phi}$ can be derived from (3), (4) and (5) as:

$$p(\boldsymbol{\theta}|\boldsymbol{\phi}, \boldsymbol{\omega}^2, \beta) = (2\pi/\beta)^{-N_\theta/2}|\mathbf{H}^T\mathbf{H}|^{1/2}\exp\left\{-\frac{\beta}{2}(\boldsymbol{\theta} - (\mathbf{H}^T\mathbf{H})^{-1}\mathbf{H}^T\mathbf{b})^T\mathbf{H}^T\mathbf{H}(\boldsymbol{\theta} - (\mathbf{H}^T\mathbf{H})^{-1}\mathbf{H}^T\mathbf{b})\right\}$$

$$= \mathcal{N}(\boldsymbol{\theta}|(\mathbf{H}^T\mathbf{H})^{-1}\mathbf{H}^T\mathbf{b}, (\beta\mathbf{H}^T\mathbf{H})^{-1}) \tag{6a}$$

$$p(\boldsymbol{\omega}^2|\boldsymbol{\theta}, \boldsymbol{\phi}, \beta) = (2\pi/\beta)^{-N_m/2}|\mathbf{G}^T\mathbf{G}|^{1/2}\exp\left\{-\frac{\beta}{2}(\boldsymbol{\omega}^2 - (\mathbf{G}^T\mathbf{G})^{-1}\mathbf{G}^T\mathbf{c})^T\mathbf{G}^T\mathbf{G}(\boldsymbol{\omega}^2 - (\mathbf{G}^T\mathbf{G})^{-1}\mathbf{G}^T\mathbf{c})\right\}$$

$$= \mathcal{N}(\boldsymbol{\omega}^2|(\mathbf{G}^T\mathbf{G})^{-1}\mathbf{G}^T\mathbf{c}, (\beta\mathbf{G}^T\mathbf{G})^{-1}) \tag{6b}$$

$$p(\boldsymbol{\phi}|\boldsymbol{\omega}^2, \boldsymbol{\theta}, \beta) = (2\pi/\beta)^{-N_d N_m/2}|\mathbf{F}^T\mathbf{F}|^{1/2}\exp\left\{-\frac{\beta}{2}\boldsymbol{\phi}^T\mathbf{F}^T\mathbf{F}\boldsymbol{\phi}\right\} = \mathcal{N}(\boldsymbol{\phi}|\mathbf{0}, (\beta\mathbf{F}^T\mathbf{F})^{-1}) \tag{6c}$$

where:

$$\mathbf{H} = \begin{bmatrix} \mathbf{K}_1\boldsymbol{\phi}_1 & \cdots & \mathbf{K}_{N_\theta}\boldsymbol{\phi}_1 \\ \vdots & \ddots & \vdots \\ \mathbf{K}_1\boldsymbol{\phi}_{N_m} & \cdots & \mathbf{K}_{N_\theta}\boldsymbol{\phi}_{N_m} \end{bmatrix}_{N_m N_d \times N_\theta}, \quad \mathbf{b} = \begin{bmatrix} (\omega_1^2\,\mathbf{M} - \mathbf{K}_0)\boldsymbol{\phi}_1 \\ \vdots \\ (\omega_{N_m}^2\mathbf{M} - \mathbf{K}_0)\boldsymbol{\phi}_{N_m} \end{bmatrix}_{N_m N_d \times 1}, \tag{7a}$$

$$\mathbf{G} = \begin{bmatrix} \mathbf{M}\boldsymbol{\phi}_1 & \cdots & 0 \\ \vdots & \ddots & \vdots \\ 0 & \cdots & \mathbf{M}\boldsymbol{\phi}_{N_m} \end{bmatrix}_{N_m N_d \times N_m}, \quad \mathbf{c} = \begin{bmatrix} \mathbf{K}(\boldsymbol{\theta})\boldsymbol{\phi}_1 \\ \vdots \\ \mathbf{K}(\boldsymbol{\theta})\boldsymbol{\phi}_{N_m} \end{bmatrix}_{N_m N_d \times 1}, \tag{7b}$$

$$\mathbf{F} = \begin{bmatrix} \mathbf{K}(\boldsymbol{\theta}) - \omega_1^2\mathbf{M} & \cdots & \mathbf{0} \\ \vdots & \ddots & \vdots \\ \mathbf{0} & \cdots & \mathbf{K}(\boldsymbol{\theta}) - \omega_{N_m}^2\mathbf{M} \end{bmatrix}_{N_m N_d \times N_m N_d} \tag{7c}$$

For the equation-error precision parameter $\beta$, we take the widely used exponential prior PDF:

$$p(\beta|b_0) = \text{Exp}(\beta|b_0) = b_0 \exp(-b_0\beta) \tag{8}$$

which is the maximum entropy prior for $\beta$ with the mean constraint $\mathbf{E}(\beta|b_0) = 1/b_0$. The hyper-prior for $b_0$ is taken as a locally non-informative one (i.e. uniform over a sufficiently large interval $(0, b_{0,max})$).

*2.3.2. Likelihood function for system modal parameters and structural stiffness scaling parameters*

Following [15] and using the Principle of Maximum Information Entropy again, the combined prediction errors and measurement errors for the system modal parameters $\boldsymbol{\phi}$ and $\boldsymbol{\omega}^2$ are modeled independently as zero-mean Gaussian variables with unknown variances, so the likelihood functions for $\boldsymbol{\phi}$ and $\boldsymbol{\omega}^2$ are given by:

$$p(\widehat{\boldsymbol{\psi}}|\boldsymbol{\phi}, \eta) = \mathcal{N}(\widehat{\boldsymbol{\psi}}|\boldsymbol{\Gamma}\boldsymbol{\phi}, \eta\mathbf{I}_{N_o N_s N_m}) \tag{9}$$

$$p(\widehat{\boldsymbol{\omega}}^2|\boldsymbol{\omega}^2, \rho) = \mathcal{N}(\widehat{\boldsymbol{\omega}}^2|\mathbf{T}\boldsymbol{\omega}^2, \rho\mathbf{I}_{N_s N_m}) \tag{10}$$

where $\boldsymbol{\Gamma} \in \mathbb{R}^{N_o N_s N_m \times N_d N_m}$ with "1s" and "0s" picks the observed degrees of freedom in the "measured" mode shape data set from the full system mode shapes $\boldsymbol{\phi}$; $\mathbf{T} = [\mathbf{I}_{N_m}, \dots, \mathbf{I}_{N_m}]^T \in \mathbb{R}^{N_s N_m \times N_m}$ is the matrix which connects the vector of $N_s$ sets of $N_m$ identified natural frequencies $\widehat{\boldsymbol{\omega}}^2$ and the $N_m$ system natural frequencies $\boldsymbol{\omega}^2$. Parameters $\eta$ and $\rho$ are prescribed variances for the predictions of the identified mode shapes $\widehat{\boldsymbol{\psi}}$ and natural frequencies $\widehat{\boldsymbol{\omega}}^2$ from $\boldsymbol{\phi}$ and $\boldsymbol{\omega}^2$, respectively. The hyper-priors for $\eta$ and $\rho$ are taken as locally non-informative ones.

For the structural stiffness scaling parameters $\boldsymbol{\theta}$, we assume there is a "calibration" value $\widehat{\boldsymbol{\theta}}$ available, which is derived theoretically (e.g. from a finite-element structure model) or is identified from previous experimental data, and we want to impose sparsity on the change $\Delta\boldsymbol{\theta} = \boldsymbol{\theta} - \widehat{\boldsymbol{\theta}}$ to reduce ill-conditioning. For example, in structure health monitoring, this could be motivated by the fact that damage-induced stiffness changes typically will be localized in a small number of substructures, such as the connections of steel members where local buckling or weld fracture can occur.

For the case where the user wants to induce sparseness in the change $\Delta\boldsymbol{\theta} = \boldsymbol{\theta} - \widehat{\boldsymbol{\theta}}$, we take a Gaussian pseudo-likelihood function:

$$p(\widehat{\boldsymbol{\theta}}|\boldsymbol{\theta}, \boldsymbol{\alpha}) = \prod_{j=1}^{N_\theta} \mathcal{N}(\hat{\theta}_j|\theta_j, \alpha_j) = \mathcal{N}(\widehat{\boldsymbol{\theta}}|\boldsymbol{\theta}, \mathbf{A}) \tag{11}$$

where $\mathbf{A} = \text{diag}(\alpha_1, \dots, \alpha_{N_\theta})$. This probability model corresponds to the maximum entropy PDF subjected to only the first two moment constraints of zero mean and independent variances $\alpha_j$ for each component $\Delta\theta_j$ of $\Delta\boldsymbol{\theta}$. It gives a measure of the plausibility of the calibration value $\widehat{\boldsymbol{\theta}}$ when the structural model is specified by the parameter vector $\boldsymbol{\theta}$. the parameter vector $\boldsymbol{\alpha}$ is learned from the modal data using a locally non-informative prior on $\boldsymbol{\alpha}$. This approach to induce sparseness is based on the idea of using the Automatic Relevance Determination Gaussian prior in sparse Bayesian learning [17].

On the other hand, if a sparse stiffness change is not desired, then the pseudo-likelihood function in (11) is not required for the stiffness scaling parameter $\boldsymbol{\theta}$ and only the likelihood function $p(\hat{\boldsymbol{\psi}}, \hat{\boldsymbol{\omega}}^2|\boldsymbol{\phi}, \boldsymbol{\omega}^2, \boldsymbol{\theta}) = p(\hat{\boldsymbol{\psi}}|\boldsymbol{\phi})p(\hat{\boldsymbol{\omega}}^2|\boldsymbol{\omega}^2)$ is used for Bayesian model updating. Theoretically, this choice corresponds to all $\alpha_j \to \infty$ in (11), so $\hat{\boldsymbol{\theta}}$ is irrelevant. It would be appropriate, for example, if: (1) sufficient modal data are available to allow a non-sparse set of stiffness changes to be reliably inferred if needed; or (2) there is no reliable calibration value $\hat{\boldsymbol{\theta}}$ available to impose sparsity on the change $\Delta\boldsymbol{\theta}$.

*2.3.3. Full posterior PDF*

To simplify the notation, we denote the pseudo-data used for structural model updating as $\boldsymbol{\mathcal{D}} = \left[\hat{\boldsymbol{\psi}}^T, (\hat{\boldsymbol{\omega}}^2)^T, \hat{\boldsymbol{\theta}}^T\right]^T$ if the pseudo-likelihood function in (11) is assigned, otherwise $\hat{\boldsymbol{\theta}}$ is dropped from $\boldsymbol{\mathcal{D}}$. For system identification, our goal is to draw samples that characterize $p(\boldsymbol{\theta}|\boldsymbol{\mathcal{D}})$, the marginal posterior PDF of the stiffness scaling parameters. These samples are obtained by sampling from the joint posterior PDF $p(\boldsymbol{\phi}, \boldsymbol{\omega}^2, \boldsymbol{\theta}, \beta|\boldsymbol{\mathcal{D}})$ and selecting the $\boldsymbol{\theta}$ components. Based on the hierarchical model represented in Figure 1, this posterior PDF can be calculated by marginalizing over the parameters $\eta, \rho, \boldsymbol{\alpha}$ and $b_0$ in Bayes' theorem as follows:

$$p(\boldsymbol{\omega}^2, \boldsymbol{\phi}, \boldsymbol{\theta}, \beta|\boldsymbol{\mathcal{D}}) = p(\hat{\boldsymbol{\omega}}^2|\boldsymbol{\omega}^2)p(\hat{\boldsymbol{\psi}}|\boldsymbol{\phi})p(\hat{\boldsymbol{\theta}}|\boldsymbol{\theta})p(\boldsymbol{\omega}^2, \boldsymbol{\phi}, \boldsymbol{\theta}|\beta)p(\beta)/p(\hat{\boldsymbol{\omega}}^2, \hat{\boldsymbol{\psi}}, \hat{\boldsymbol{\theta}})$$

$$= \int p(\hat{\boldsymbol{\omega}}^2|\boldsymbol{\omega}^2, \rho)p(\hat{\boldsymbol{\psi}}|\boldsymbol{\phi}, \eta)p(\hat{\boldsymbol{\theta}}|\boldsymbol{\theta}, \boldsymbol{\alpha})p(\boldsymbol{\omega}^2, \boldsymbol{\phi}, \boldsymbol{\theta}|\beta)p(\rho)p(\eta)p(\boldsymbol{\alpha})p(\beta|b_0)p(b_0)d\rho d\eta d\boldsymbol{\alpha} db_0/p(\hat{\boldsymbol{\omega}}^2, \hat{\boldsymbol{\psi}}, \hat{\boldsymbol{\theta}}) \quad (12)$$

where PDFs $p(\hat{\boldsymbol{\omega}}^2|\boldsymbol{\omega}^2, \rho), p(\hat{\boldsymbol{\psi}}|\boldsymbol{\phi}, \eta), p(\hat{\boldsymbol{\theta}}|\boldsymbol{\theta}, \boldsymbol{\alpha}), p(\boldsymbol{\omega}^2, \boldsymbol{\phi}, \boldsymbol{\theta}|\beta)$ and $p(\beta|b_0)$are defined in (10), (9), (11), (3) and (8), respectively. The prior PDFs for $\rho, \eta, \alpha_j$ and $b_0$ are taken as locally non-informative ones over large open intervals that start at zero. The resulting expression is intractable because the high-dimensional normalizing integral $p(\hat{\boldsymbol{\omega}}^2, \hat{\boldsymbol{\psi}}, \hat{\boldsymbol{\theta}})$ cannot be computed analytically. Instead, we implement GS to draw posterior samples from $p(\boldsymbol{\phi}, \boldsymbol{\omega}^2, \boldsymbol{\theta}, \beta|\boldsymbol{\mathcal{D}})$ by decomposing the whole model parameter vector into the four groups $\{\boldsymbol{\phi}, \boldsymbol{\omega}^2, \boldsymbol{\theta}, \beta\}$ and repeatedly sampling from one parameter group conditional on the other three groups and the available data. We show in Appendix A that the required conditional posterior PDFs to implement GS can be established analytically. To improve the efficiency of the GS procedure, we also develop an algorithm where we integrate out the equation-error precision parameter $\beta$ and repeatedly draw conditional samples from only three groups of parameters $\{\boldsymbol{\phi}, \boldsymbol{\omega}^2, \boldsymbol{\theta}\}$, to finally sample the posterior PDF $p(\boldsymbol{\phi}, \boldsymbol{\omega}^2, \boldsymbol{\theta}|\boldsymbol{\mathcal{D}})$. The conditional posterior PDFs to implement GS in this case are derived in Appendix B. In the next section, the full conditional posterior PDFs for each parameter group are derived.

## 3. GIBBS SAMPLING ALGORITHM FOR BAYESIAN SYSTEM IDENTIFICATION

### *3.1 Full conditional posterior PDFs without marginalizing over equation-error precision $\beta$*

*3.1.1 Conditional posterior PDF for $\boldsymbol{\phi}$*

Using the general result in Appendix A, the conditional posterior PDF over parameter vector $\boldsymbol{\phi}$ is obtained as:

$$p(\boldsymbol{\phi}|\boldsymbol{\mathcal{D}},\boldsymbol{\omega}^2,\boldsymbol{\theta},\beta) = \int p(\boldsymbol{\phi}|\boldsymbol{\mathcal{D}},\boldsymbol{\omega}^2,\boldsymbol{\theta},\beta,\eta)p(\eta|\boldsymbol{\mathcal{D}},\boldsymbol{\omega}^2,\boldsymbol{\theta},\beta)d\eta \approx p(\boldsymbol{\phi}|\boldsymbol{\mathcal{D}},\boldsymbol{\omega}^2,\boldsymbol{\theta},\beta,\tilde{\eta})$$
$$\propto p(\widehat{\boldsymbol{\psi}}|\boldsymbol{\phi},\tilde{\eta})p(\boldsymbol{\phi}|\boldsymbol{\omega}^2,\boldsymbol{\theta},\beta) = \mathcal{N}(\boldsymbol{\phi}|\boldsymbol{\mu}_{\boldsymbol{\phi}},\boldsymbol{\Sigma}_{\boldsymbol{\phi}}) \tag{13}$$

where：

$$\boldsymbol{\mu}_{\boldsymbol{\phi}} = \tilde{\eta}^{-1}\boldsymbol{\Sigma}_{\boldsymbol{\phi}}\boldsymbol{\Gamma}^T\widehat{\boldsymbol{\psi}} \tag{14a}$$

$$\boldsymbol{\Sigma}_{\boldsymbol{\phi}} = (\beta\mathbf{F}^T\mathbf{F} + \tilde{\eta}^{-1}\boldsymbol{\Gamma}^T\boldsymbol{\Gamma})^{-1} \tag{14b}$$

$$\tilde{\eta} = \frac{1}{N_o N_s N_m}\left[\mathrm{tr}(\boldsymbol{\Sigma}_{\boldsymbol{\phi}}\boldsymbol{\Gamma}^T\boldsymbol{\Gamma}) + \left\|\widehat{\boldsymbol{\psi}} - \boldsymbol{\Gamma}\boldsymbol{\mu}_{\boldsymbol{\phi}}\right\|^2\right] \tag{15}$$

where $\mathrm{tr}(\cdot)$ denotes the trace of a matrix and $\mathbf{F}(\boldsymbol{\omega}^2,\boldsymbol{\theta})$ is given in (7c). For given $\boldsymbol{\omega}^2$, $\boldsymbol{\theta}$ and $\beta$, both the posterior mean $\boldsymbol{\mu}_{\boldsymbol{\phi}}$ and covariance matrix $\boldsymbol{\Sigma}_{\boldsymbol{\phi}}$ depend on the MAP value $\tilde{\eta}$, and so an iterative method, cycling over (14a), (14b) and (15) until convergence, is required.

*3.1.2 Conditional posterior PDF for* $\boldsymbol{\omega}^2$

According to the general result in Appendix A, the conditional posterior PDF for vector $\boldsymbol{\omega}^2$ is given by:

$$p(\boldsymbol{\omega}^2|\boldsymbol{\mathcal{D}},\boldsymbol{\phi},\boldsymbol{\theta},\beta) \approx p(\boldsymbol{\omega}^2|\boldsymbol{\mathcal{D}},\boldsymbol{\phi},\boldsymbol{\theta},\beta,\tilde{\rho}) \propto p(\widehat{\boldsymbol{\omega}}^2|\boldsymbol{\omega}^2,\tilde{\rho})p(\boldsymbol{\omega}^2|\boldsymbol{\phi},\boldsymbol{\theta},\beta) = \mathcal{N}(\boldsymbol{\omega}^2|\boldsymbol{\mu}_{\boldsymbol{\omega}^2},\boldsymbol{\Sigma}_{\boldsymbol{\omega}^2}) \tag{16}$$

where:

$$\boldsymbol{\mu}_{\boldsymbol{\omega}^2} = \boldsymbol{\Sigma}_{\boldsymbol{\omega}^2}(\beta\mathbf{G}^T\mathbf{c} + \tilde{\rho}^{-1}\mathbf{T}^T\widehat{\boldsymbol{\omega}}^2) \tag{17a}$$

$$\boldsymbol{\Sigma}_{\boldsymbol{\omega}^2} = (\tilde{\rho}^{-1}\mathbf{T}^T\mathbf{T} + \beta\mathbf{G}^T\mathbf{G})^{-1} \tag{17b}$$

$$\tilde{\rho} = \frac{1}{N_s N_m}[\mathrm{tr}(\boldsymbol{\Sigma}_{\boldsymbol{\omega}^2}\mathbf{T}^T\mathbf{T}) + \|\widehat{\boldsymbol{\omega}}^2 - \mathbf{T}\boldsymbol{\mu}_{\boldsymbol{\omega}^2}\|^2] \tag{18}$$

$\mathbf{G}(\boldsymbol{\phi})$ and $\mathbf{c}(\boldsymbol{\phi},\boldsymbol{\theta})$ are given in (7b). An iterative method is then used to determine $\boldsymbol{\mu}_{\boldsymbol{\omega}^2}$, $\boldsymbol{\Sigma}_{\boldsymbol{\omega}^2}$ and the MAP value $\tilde{\rho}$ by cycling over (17a), (17b) and (18).

*3.1.3 Conditional posterior PDF for* $\boldsymbol{\theta}$

From Appendix A, the conditional posterior PDF for $\boldsymbol{\theta}$ is derived as:

$$p(\boldsymbol{\theta}|\boldsymbol{\mathcal{D}},\boldsymbol{\phi},\boldsymbol{\omega}^2,\beta) \approx p(\boldsymbol{\theta}|\boldsymbol{\mathcal{D}},\boldsymbol{\phi},\boldsymbol{\omega}^2,\beta,\tilde{\boldsymbol{\alpha}}) \propto p(\widehat{\boldsymbol{\theta}}|\boldsymbol{\theta},\tilde{\boldsymbol{\alpha}})p(\boldsymbol{\theta}|\boldsymbol{\phi},\boldsymbol{\omega}^2,\beta) = \mathcal{N}(\boldsymbol{\theta}|\boldsymbol{\mu}_{\boldsymbol{\theta}},\boldsymbol{\Sigma}_{\boldsymbol{\theta}}) \tag{19}$$

where

$$\boldsymbol{\mu}_{\boldsymbol{\theta}} = \boldsymbol{\Sigma}_{\boldsymbol{\theta}}(\beta\mathbf{H}^T\mathbf{b} + \widetilde{\mathbf{A}}^{-1}\widehat{\boldsymbol{\theta}}) \tag{20a}$$

$$\boldsymbol{\Sigma}_{\boldsymbol{\theta}} = (\beta\mathbf{H}^T\mathbf{H} + \widetilde{\mathbf{A}}^{-1})^{-1} \tag{20b}$$

and $\mathbf{H}(\boldsymbol{\phi})$and $\mathbf{b}(\boldsymbol{\phi},\boldsymbol{\omega}^2)$ are given in (7a) and $\widetilde{\mathbf{A}} = \mathrm{diag}(\tilde{\alpha}_1,\dots,\tilde{\alpha}_{N_\theta})$.

For the case where a sparse change $\Delta\boldsymbol{\theta} = \boldsymbol{\theta} - \widehat{\boldsymbol{\theta}}$ is desired, $\tilde{\boldsymbol{\alpha}} = \arg\max p(\boldsymbol{\alpha}|\mathbf{y},\boldsymbol{\phi},\boldsymbol{\omega}^2,\beta)$ is given by using (A12), so for each $j = 1,\dots,N_{\boldsymbol{\theta}}$:

$$\tilde{\alpha}_j = (\boldsymbol{\Sigma}_{\boldsymbol{\theta}})_{jj} + \left(\hat{\boldsymbol{\theta}} - \boldsymbol{\mu}_{\boldsymbol{\theta}}\right)_j^2 \tag{21}$$

An iterative method is then used to determine $\boldsymbol{\mu}_{\boldsymbol{\theta}}$, $\boldsymbol{\Sigma}_{\boldsymbol{\theta}}$ and $\tilde{\boldsymbol{\alpha}}$ by cycling over (20a), (20b) and (21). It is expected that many of the $\tilde{\alpha}_j$ will approach zero during the optimization, which implies from (11) that the corresponding $\Delta\theta_j = \theta_j - \hat{\theta}_j$ will have negligibly small values [15]). This is a similar procedure to the original sparse Bayesian learning where redundant or irrelevant features are pruned away to produce a sparse explanatory subset by learning the ARD prior variances [17].

If the user does not wish to impose sparseness on the changes in $\boldsymbol{\theta}$ during Bayesian updating based on modal data, the conditional posterior PDF for $\boldsymbol{\theta}$ is obtained by simply setting $\tilde{\alpha}_j \to \infty$ in (19,20) and so the terms in (20a,b) involving $\tilde{\mathbf{A}}$ are excluded and (21) is not relevant.

*3.1.4 Conditional posterior PDF for $\beta$*

The conditional posterior PDF for $\beta$ is derived as:

$$p(\beta|\boldsymbol{\mathcal{D}}, \boldsymbol{\phi}, \boldsymbol{\omega}^2, \boldsymbol{\theta}) = \int p(\beta|\boldsymbol{\mathcal{D}}, \boldsymbol{\phi}, \boldsymbol{\omega}^2, \boldsymbol{\theta}, b_0)\, p(b_0|\boldsymbol{\mathcal{D}}, \boldsymbol{\phi}, \boldsymbol{\omega}^2, \boldsymbol{\theta}) db_0$$

$$\approx p\left(\beta|\boldsymbol{\mathcal{D}}, \boldsymbol{\phi}, \boldsymbol{\omega}^2, \boldsymbol{\theta}, \tilde{b}_0\right) \propto p(\boldsymbol{\mathcal{D}}, \boldsymbol{\phi}, \boldsymbol{\omega}^2, \boldsymbol{\theta}|\beta)\, p\left(\beta|\tilde{b}_0\right) = p(\boldsymbol{\mathcal{D}}|\boldsymbol{\phi}, \boldsymbol{\omega}^2, \boldsymbol{\theta})\, p(\boldsymbol{\phi}, \boldsymbol{\omega}^2, \boldsymbol{\theta}|\beta) p\left(\beta|\tilde{b}_0\right)$$

$$\propto p(\boldsymbol{\phi}, \boldsymbol{\omega}^2, \boldsymbol{\theta}|\beta) p\left(\beta|\tilde{b}_0\right) \propto \mathrm{Gamma}(\beta|a_0', b_0') \tag{22}$$

where $\tilde{b}_0 = \arg\max p(b_0|\boldsymbol{\mathcal{D}}, \boldsymbol{\phi}, \boldsymbol{\omega}^2, \boldsymbol{\theta})$ and the shape parameter $a_0'$ and rate parameter $b_0'$ for the posterior gamma distribution on $\beta$ are given by:

$$a_0' = 1 + N_m N_d/2 \tag{23a}$$

$$b_0' = \tilde{b}_0 + \sum_{i=1}^{N_m} \left\| \left(\mathbf{K}(\boldsymbol{\theta}) - \omega_i^2 \mathbf{M}\right) \boldsymbol{\phi}_i \right\|^2 / 2. \tag{23b}$$

Since $b_0$ has a uniform prior, the posterior PDF for $b_0$ is derived as:

$$p(b_0|\boldsymbol{\mathcal{D}}, \boldsymbol{\phi}, \boldsymbol{\omega}^2, \boldsymbol{\theta}) \propto p(\boldsymbol{\mathcal{D}}, \boldsymbol{\phi}, \boldsymbol{\omega}^2, \boldsymbol{\theta}|b_0) p(b_0) \propto p(\boldsymbol{\mathcal{D}}, \boldsymbol{\phi}, \boldsymbol{\omega}^2, \boldsymbol{\theta}|b_0) = p(\boldsymbol{\mathcal{D}}|\boldsymbol{\phi}, \boldsymbol{\omega}^2, \boldsymbol{\theta})\, p(\boldsymbol{\phi}, \boldsymbol{\omega}^2, \boldsymbol{\theta}|b_0)$$

$$\propto p(\boldsymbol{\phi}, \boldsymbol{\omega}^2, \boldsymbol{\theta}|b_0) = \int p(\boldsymbol{\phi}, \boldsymbol{\omega}^2, \boldsymbol{\theta}|\beta)\, p(\beta|b_0)\, d\beta$$

$$= \frac{\Gamma(1 + N_m N_d/2) {b_0}^{-N_m N_d/2}}{(2\pi)^{N_m N_d/2}} \left\{1 + \frac{1}{2b_0}\left(\sum_{i=1}^{N_m} \left\| \left(\mathbf{K}(\boldsymbol{\theta}) - \omega_i^2 \mathbf{M}\right) \boldsymbol{\phi}_i \right\|^2\right)\right\}^{-N_m N_d/2 - 1} \tag{24}$$

By direct differentiation of the logarithm of (24) with respect to $b_0$, the MAP estimate of $b_0$ is given by:

$$\tilde{b}_0 = \frac{1}{N_m N_d} \sum_{i=1}^{N_m} \left\| \left(\mathbf{K}(\boldsymbol{\theta}) - \omega_i^2 \mathbf{M}\right) \boldsymbol{\phi}_i \right\|^2 \tag{25}$$

### *3.2 Full conditional posterior PDFs when marginalizing over equation-error precision* $\beta$

When marginalizing over the equation error precision parameter $\beta$, GS samples are drawn from the three conditional posterior PDFs $p(\boldsymbol{\phi}|\boldsymbol{\mathcal{D}}, \boldsymbol{\omega}^2, \boldsymbol{\theta}), p(\boldsymbol{\omega}^2|\boldsymbol{\mathcal{D}}, \boldsymbol{\phi}, \boldsymbol{\theta})$ and $p(\boldsymbol{\theta}|\boldsymbol{\mathcal{D}}, \boldsymbol{\phi}, \boldsymbol{\omega}^2)$, which are derived as follows.

#### *3.2.1 Conditional posterior PDF for* $\boldsymbol{\phi}$

Following the strategy in Appendix B with the substitution $\tau = \beta\eta$, the conditional posterior PDF for $\boldsymbol{\phi}$:

$$p(\boldsymbol{\phi}|\boldsymbol{\mathcal{D}}, \boldsymbol{\omega}^2, \boldsymbol{\theta}) = \int p(\boldsymbol{\phi}|\boldsymbol{\mathcal{D}}, \boldsymbol{\omega}^2, \boldsymbol{\theta}, \beta)\, p(\beta|\boldsymbol{\mathcal{D}}, \boldsymbol{\omega}^2, \boldsymbol{\theta}) d\beta = \mathrm{St}\left(\boldsymbol{\phi}|\boldsymbol{\mu}_{\boldsymbol{\phi}}, \frac{b_{\boldsymbol{\phi}}^{(1)}}{a_{\boldsymbol{\phi}}^{(1)}}\boldsymbol{\Lambda}_{\boldsymbol{\phi}}, 2a_0^{(1)}\right) \tag{26}$$

where

$$\boldsymbol{\mu}_{\boldsymbol{\phi}} = \tilde{\tau}^{-1}\boldsymbol{\Lambda}_{\boldsymbol{\phi}}^{-1}\boldsymbol{\Gamma}^T\widehat{\boldsymbol{\psi}} \tag{27a}$$

$$\boldsymbol{\Lambda}_{\boldsymbol{\phi}} = \tilde{\tau}^{-1}\boldsymbol{\Gamma}^T\boldsymbol{\Gamma} + \mathbf{F}^T\mathbf{F} \tag{27b}$$

$$a_{\boldsymbol{\phi}}^{(1)} = N_o N_s N_m/2 + 1 \tag{27c}$$

$$b_{\boldsymbol{\phi}}^{(1)} = \tilde{\tau}^{-1}\left\|\widehat{\boldsymbol{\psi}} - \boldsymbol{\Gamma}\boldsymbol{\mu}_{\boldsymbol{\phi}}\right\|^2/2 + \left\|\mathbf{F}\boldsymbol{\mu}_{\boldsymbol{\phi}}\right\|^2/2 + \tilde{b}_{0,\boldsymbol{\phi}} \tag{27d}$$

$$\tilde{b}_{0,\boldsymbol{\phi}} = \frac{1}{N_o N_s N_m}\left(\tilde{\tau}^{-1}\left\|\widehat{\boldsymbol{\psi}} - \boldsymbol{\Gamma}\boldsymbol{\mu}_{\boldsymbol{\phi}}\right\|^2 + \left\|\mathbf{F}\boldsymbol{\mu}_{\boldsymbol{\phi}}\right\|^2\right) \tag{27e}$$

$$\tilde{\tau} = \frac{1}{N_o N_s N_m}\left[\mathrm{tr}\left(\boldsymbol{\Lambda}_{\boldsymbol{\phi}}^{-1}\boldsymbol{\Gamma}^T\boldsymbol{\Gamma}\right) + \frac{a_{\boldsymbol{\phi}}^{(1)}\left\|\widehat{\boldsymbol{\psi}} - \boldsymbol{\Gamma}\boldsymbol{\mu}_{\boldsymbol{\phi}}\right\|^2}{b_{\boldsymbol{\phi}}^{(1)}}\right] \tag{28}$$

In (28), both the posterior mean $\boldsymbol{\mu}_{\boldsymbol{\phi}}$ and matrix $\boldsymbol{\Lambda}_{\boldsymbol{\phi}}$ depend on the MAP value $\tilde{\tau}$, and so they are obtained by an iterative method using (27a), (27b), (27c), (27d), (27e) and (28).

#### *3.2.2 Conditional posterior PDF for* $\boldsymbol{\omega}^2$

Following the same strategy as in Section 3.2.1 but with the substitution $\upsilon = \beta\rho$, the conditional posterior PDF for $\boldsymbol{\omega}^2$ is given by:

$$p(\boldsymbol{\omega}^2|\boldsymbol{\mathcal{D}}, \boldsymbol{\phi}, \boldsymbol{\theta}) = \int p(\boldsymbol{\omega}^2|\boldsymbol{\mathcal{D}}, \boldsymbol{\phi}, \boldsymbol{\theta}, \beta)\, p(\beta|\boldsymbol{\mathcal{D}}, \boldsymbol{\phi}, \boldsymbol{\theta}) d\beta = \mathrm{St}\left(\boldsymbol{\omega}^2|\boldsymbol{\mu}_{\boldsymbol{\omega}^2}, \frac{b_{\boldsymbol{\omega}^2}^{(1)}}{a_{\boldsymbol{\omega}^2}^{(1)}}\boldsymbol{\Lambda}_{\boldsymbol{\omega}^2}, 2a_{\boldsymbol{\omega}^2}^{(1)}\right) \tag{29}$$

where

$$\boldsymbol{\mu}_{\boldsymbol{\omega}^2} = \boldsymbol{\Lambda}_{\boldsymbol{\omega}^2}(\mathbf{G}^T\mathbf{c} + \tilde{\upsilon}^{-1}\mathbf{T}^T\widehat{\boldsymbol{\omega}}^2) \tag{30a}$$

$$\boldsymbol{\Lambda}_{\boldsymbol{\omega}^2} = \tilde{\upsilon}^{-1}\mathbf{T}^T\mathbf{T} + \mathbf{G}^T\mathbf{G} \tag{30b}$$

$$a_{\boldsymbol{\omega}^2}^{(1)} = (N_d N_m + N_s N_m - N_m)/2 + 1 \tag{30c}$$

$$b_{\boldsymbol{\omega}^2}^{(1)} = \tilde{\upsilon}^{-1}\,\|\widehat{\boldsymbol{\omega}}^2 - \mathbf{T}\boldsymbol{\mu}_{\boldsymbol{\omega}^2}\|^2/2 + \|\mathbf{G}\boldsymbol{\mu}_{\boldsymbol{\omega}^2} - \mathbf{c}\|^2/2 + \tilde{b}_{0,\boldsymbol{\omega}^2} \tag{30d}$$

$$\tilde{b}_{0,\boldsymbol{\omega}^2} = \frac{1}{N_d N_m + N_s N_m - N_m}(\tilde{\upsilon}^{-1}\|\widehat{\boldsymbol{\omega}}^2 - \mathbf{T}\boldsymbol{\mu}_{\boldsymbol{\omega}^2}\|^2 + \|\mathbf{G}\boldsymbol{\mu}_{\boldsymbol{\omega}^2} - \mathbf{c}\|^2) \tag{30e}$$

$$\tilde{\upsilon} = \frac{1}{N_s N_m}\left[\mathrm{tr}\left(\boldsymbol{\Lambda}_{\boldsymbol{\omega}^2}^{-1}\mathbf{T}^T\mathbf{T}\right) + \frac{a_{\boldsymbol{\omega}^2}^{(1)}\|\widehat{\boldsymbol{\omega}}^2 - \mathbf{L}\boldsymbol{\mu}_{\boldsymbol{\omega}^2}\|^2}{b_{\boldsymbol{\omega}^2}^{(1)}}\right] \tag{31}$$

These parameter values, including the MAP value $\tilde{\upsilon}$, are obtained by iterating over (30a), (30b), (30c), (30d), (30e) and (31).

### *3.2.3 Conditional posterior PDF for* $\boldsymbol{\theta}$

Similarly, with the substitution of $\boldsymbol{\gamma} = \beta\boldsymbol{\alpha}$, the conditional posterior PDF for $\boldsymbol{\theta}$ is derived as:

$$p(\boldsymbol{\theta}|\boldsymbol{\mathcal{D}}, \boldsymbol{\phi}, \boldsymbol{\omega}^2) = \int p(\boldsymbol{\theta}|\boldsymbol{\mathcal{D}}, \boldsymbol{\phi}, \boldsymbol{\omega}^2, \beta)\, p(\beta|\boldsymbol{\mathcal{D}}, \boldsymbol{\phi}, \boldsymbol{\omega}^2) d\beta = \mathrm{St}\left(\boldsymbol{\theta}|\boldsymbol{\mu}_{\boldsymbol{\theta}}, \frac{b_{\boldsymbol{\theta}}^{(1)}}{a_{\boldsymbol{\theta}}^{(1)}}\boldsymbol{\Lambda}_{\boldsymbol{\theta}}, 2a_{\boldsymbol{\theta}}^{(1)}\right) \tag{32}$$

where

$$\boldsymbol{\mu}_{\boldsymbol{\theta}} = \boldsymbol{\Lambda}_{\boldsymbol{\theta}}^{-1}\left(\mathbf{H}^T\mathbf{b} + \tilde{\boldsymbol{\gamma}}\widehat{\boldsymbol{\theta}}\right) \tag{33a}$$

$$\boldsymbol{\Lambda}_{\boldsymbol{\theta}} = \mathrm{diag}(\tilde{\boldsymbol{\gamma}}^{-1}) + \mathbf{H}^T\mathbf{H} \tag{33b}$$

$$a_{\boldsymbol{\theta}}^{(1)} = N_d N_m/2 + 1 \tag{33c}$$

$$b_{\boldsymbol{\theta}}^{(1)} = \left(\widehat{\boldsymbol{\theta}} - \boldsymbol{\mu}_{\boldsymbol{\theta}}\right)^T \mathrm{diag}(\tilde{\boldsymbol{\gamma}}^{-1})\left(\widehat{\boldsymbol{\theta}} - \boldsymbol{\mu}_{\boldsymbol{\theta}}\right)/2 + ||\mathbf{H}\boldsymbol{\mu}_{\boldsymbol{\theta}} - \mathbf{b}||^2/2 + \tilde{b}_{0,\boldsymbol{\theta}} \tag{33d}$$

$$\tilde{b}_{0,\boldsymbol{\theta}} = \frac{1}{N_d N_m}\left(\left(\widehat{\boldsymbol{\theta}} - \boldsymbol{\mu}_{\boldsymbol{\theta}}\right)^T \mathrm{diag}(\tilde{\boldsymbol{\gamma}}^{-1})\left(\widehat{\boldsymbol{\theta}} - \boldsymbol{\mu}_{\boldsymbol{\theta}}\right) + ||\mathbf{H}\boldsymbol{\mu}_{\boldsymbol{\theta}} - \mathbf{b}||^2\right) \tag{33e}$$

and if sparseness in $\Delta\boldsymbol{\theta} = \boldsymbol{\theta} - \widehat{\boldsymbol{\theta}}$ is desired, then:

$$\tilde{\gamma}_j = \left(\boldsymbol{\Lambda}_{\theta}^{-1}\right)_{jj} + \frac{a_{\boldsymbol{\theta}}^{(1)}\left(\widehat{\boldsymbol{\theta}} - \boldsymbol{\mu}_{\boldsymbol{\theta}}\right)_j^2}{b_{\boldsymbol{\theta}}^{(1)}} \tag{34}$$

These parameter values, including the MAP value $\tilde{\gamma}_j$, are obtained by interating over (33a), (33b), (33c), (33d), (33e) and (34). Like the $\tilde{\alpha}_j$, many of the $\tilde{\gamma}_j$ tend to zero during optimization and sparseness of the stiffness change $\Delta\boldsymbol{\theta}$ is produced. As in Subsection 3.1.3, the conditional posterior PDF for $\boldsymbol{\theta}$ can be modified by taking all $\tilde{\gamma}_j \to \infty$ in (32,33) if no sparsity of the change $\Delta\boldsymbol{\theta}$ is to be imposed. Note that the denominator in (33e) should be changed from $N_d N_m$ to $(N_d N_m - N_{\boldsymbol{\theta}})$ since the size $K$ of pseudo data $\widehat{\boldsymbol{\theta}}$ in (B9) becomes zero in this case.

### ***3.3 Pseudo-code for Gibbs sampling algorithms***

The GS (Gibbs sampling) pseudo-codes are summarized as Algorithms 1 and 2 below, which are implemented by successively sampling the conditional posterior PDFs presented in Sections 3.1 and 3.2, respectively. In the algorithms, if the Markov chain created by the GS is ergodic [34], the GS samples will be finally distributed as the full posterior PDF $p(\boldsymbol{\phi}, \boldsymbol{\omega}^2, \boldsymbol{\theta}, \beta|\boldsymbol{\mathcal{D}})$ (Algorithm 1) or $p(\boldsymbol{\phi}, \boldsymbol{\omega}^2, \boldsymbol{\theta}|\boldsymbol{\mathcal{D}})$ (Algorithm 2) when the number of samples $n$ is sufficiently large. GS is ergodic from a practical point of view when the regions of high values of the full posterior PDFs $p(\boldsymbol{\phi}, \boldsymbol{\omega}^2, \boldsymbol{\theta}, \beta|\boldsymbol{\mathcal{D}})$ or $p(\boldsymbol{\phi}, \boldsymbol{\omega}^2, \boldsymbol{\theta}|\boldsymbol{\mathcal{D}})$ are effectively connected (corresponding to the model class being either globally identifiable or unidentifiable [20], which means that sampling the Markov chain fully explores its stationary state when $n$ is large, no matter how the GS algorithm is initialized. In this case, samples from the marginal distribution $p(\boldsymbol{\theta}|\boldsymbol{\mathcal{D}})$ are readily obtained by simply examining the GS samples $\overline{\boldsymbol{\theta}}^{(n)}$ for large $n$ beyond the burn-in period. It is noted that the GS algorithm using a single Markov chain is not effective for locally identifiable cases [20], where the regions of high values of the posterior PDF are well separated. Although this case is relatively rare in practice, it can be treated by parallel sampling of GS chains from multiple starting points drawn from the prior $p(\boldsymbol{\phi},\boldsymbol{\omega}^2,\boldsymbol{\theta},\beta)$ (Algorithm 1) or $p(\boldsymbol{\phi},\boldsymbol{\omega}^2,\boldsymbol{\theta})$ (Algorithm 2).

---

**Algorithm 1: Pseudo-code implementation of GS for generating $N$ samples without integrating out equation-error precision parameter $\beta$**

---

1. Initialize the samples with $\beta^{(0)} = 100$, $(\boldsymbol{\omega}^2)^{(0)} = \sum_{r=1}^{N_s} \widehat{\boldsymbol{\omega}}_r^2/N_s$ and $\boldsymbol{\theta}^{(0)} = \widehat{\boldsymbol{\theta}}$, a chosen "calibration" value for the stiffness parameter vector. Let $n = 1$.
2. **For** $n = 1$ to $N$
3. Sample the system mode shapes $\boldsymbol{\phi}^{(n)} \sim p\left(\boldsymbol{\phi}|\boldsymbol{\mathcal{D}},(\boldsymbol{\omega}^2)^{(n-1)},\boldsymbol{\theta}^{(n-1)},\beta^{(n-1)}\right)$ using (13);
4. Sample the system natural frequencies $(\boldsymbol{\omega}^2)^{(n)} \sim p\left(\boldsymbol{\omega}^2|\boldsymbol{\mathcal{D}},\boldsymbol{\phi}^{(n)},\boldsymbol{\theta}^{(n-1)},\beta^{(n-1)}\right)$ using (16);
5. Sample the equation error precision $\beta^{(n)} \sim p\left(\beta|\boldsymbol{\mathcal{D}},\boldsymbol{\phi}^{(n)},(\boldsymbol{\omega}^2)^{(n)},\boldsymbol{\theta}^{(n-1)}\right)$ using (22);
6. Sample the stiffness scaling parameters $\boldsymbol{\theta}^{(n)} \sim p\left(\boldsymbol{\theta}|\boldsymbol{\mathcal{D}},\boldsymbol{\phi}^{(n)},(\boldsymbol{\omega}^2)^{(n)},\beta^{(n)}\right)$ using (19).
7. Let $n = n + 1$
8. **End for**

Samples $\left\{\boldsymbol{\phi}^{(n)},(\boldsymbol{\omega}^2)^{(n)},\beta^{(n)},\boldsymbol{\theta}^{(n)}: n = 1,\ldots,N\right\}$ are obtained.

---

**Algorithm 2: Pseudo-code implementation of GS for generating $N$ samples when integrating out equation-error precision parameter $\beta$**

---

1. Initialize the samples with $(\boldsymbol{\omega}^2)^{(0)} = \sum_{r=1}^{N_s} \widehat{\boldsymbol{\omega}}_r^2/N_s$ and $\boldsymbol{\theta}^{(0)} = \widehat{\boldsymbol{\theta}}$, a chosen "calibration" value for the parameter vector. Let $n = 1$.
2. **For** $n = 1$ to $N$
3. Sample the system mode shapes $\boldsymbol{\phi}^{(n)} \sim p\left(\boldsymbol{\phi}|\boldsymbol{\mathcal{D}},(\boldsymbol{\omega}^2)^{(n-1)},\boldsymbol{\theta}^{(n-1)}\right)$ using (26);
4. Sample the system natural frequencies $(\boldsymbol{\omega}^2)^{(n)} \sim p\left(\boldsymbol{\omega}^2|\boldsymbol{\mathcal{D}},\boldsymbol{\phi}^{(n)},\boldsymbol{\theta}^{(n-1)}\right)$ using (29);
5. Sample the stiffness scaling parameters $\boldsymbol{\theta}^{(n)} \sim p\left(\boldsymbol{\theta}|\boldsymbol{\mathcal{D}},\boldsymbol{\phi}^{(n)},(\boldsymbol{\omega}^2)^{(n)}\right)$ using (32).
6. Let $n = n + 1$
7. **End for**

Samples $\left\{\boldsymbol{\phi}^{(n)},(\boldsymbol{\omega}^2)^{(n)},\boldsymbol{\theta}^{(n)}: n = 1,\ldots,N\right\}$ are obtained.

---

*3.3.1. Burn-in period determination*

In the implementation of GS algorithms, it is common to discard the samples during the burn-in period before the Markov chain reaches its stationary state, but this is not easy to ascertain. Following the strategy in [30] we determine the burn-in period simply by visual inspection of a plot of the Markov chain samples as they are sequentially generated. To determine how many Markov chain samples to generate after burn-in is achieved, we visually check the convergence of the plotted stiffness samples of interest.

*Remark 3.1:* One complication in Bayesian system identification using modal data is that the model for the modal parameters characterizing the modal data $\hat{\boldsymbol{\omega}}^2$ and $\hat{\boldsymbol{\psi}}$ is a nonlinear function of the structural stiffness scaling parameter vector $\boldsymbol{\theta}$. Rather than directly tackling this challenging nonlinear inverse problem, our theory is formulated in such a way that it involves a series of coupled linear−in−the−parameter problems that allow analytical construction of the conditional posterior PDFs needed for the Gibbs Sampling procedure and so it provides an efficient way to perform Bayesian system identification.

*Remark 3.2:* In our Bayesian formulation involving the Gaussian and Exponential PDFs, there is a scale invariant property due to proper learning of all of the associated hyper-parameters. If the Markov chain created by the GS is ergodic (no matter how the GS algorithm is initialized), the identification of the stiffness scaling parameters is independent of any linear scaling of both the measured mode shapes and the mass and stiffness matrices, i.e., the system identification results do not change with a change of the units for these quantities.

## 4. GIBBS SAMPLING FOR DAMAGE ASSESSMENT

### *4.1 Generating posterior samples of $\boldsymbol{\theta}$ conditional on modal data from both monitoring and calibration stages*

For damage assessment based on modal data, the Gibbs sampling algorithms in Section 3.3 are applied in two stages, *calibration* (undamaged) and *monitoring* (possibly damaged), where the pseudo-data used for the model updating of the structural model parameters $\boldsymbol{\theta}_u$ and $\boldsymbol{\theta}_d$ are $\boldsymbol{\mathcal{D}}_u = \left[\hat{\boldsymbol{\psi}}_u^T, (\hat{\boldsymbol{\omega}}_u^2)^T\right]^T$ and $\boldsymbol{\mathcal{D}}_d = \left[\hat{\boldsymbol{\psi}}_d^T, (\hat{\boldsymbol{\omega}}_d^2)^T, \boldsymbol{\theta}_u^T\right]^T$, respectively. In the case of monitoring for damage, since any damage-induced stiffness loss typically will be localized in a small number of substructures, the user may want to reduce ill-conditioning by imposing sparsity on the change $\Delta\boldsymbol{\theta} = \boldsymbol{\theta} - \boldsymbol{\theta}_u$; here, $\boldsymbol{\theta}_u$ corresponds to the (uncertain) structural stiffness scaling parameters learned during a calibration stage for a structural model of the undamaged structure. In the calibration stage, we assume that the user does not want to impose sparseness on the changes in $\boldsymbol{\theta}$ during Bayesian updating based on the modal data, although sparse change $\Delta\boldsymbol{\theta} = \boldsymbol{\theta} - \boldsymbol{\theta}_0$ could be induced if desired where $\boldsymbol{\theta}_0$ comes from a finite-element structural model, for example. Before proceeding, we note that the posterior uncertainty for $\boldsymbol{\theta}_u$ obtained from the calibration stage is controllable since it is usually the case that large amounts of time-domain vibration data can be collected from ambient vibration tests of the undamaged structure, leading to multiple sets of the identified modal parameters, because there is no urgency to rapidly detect damage. In the following, we introduce a sampling strategy to incorporate the posterior uncertainty of $\boldsymbol{\theta}_u$ in the posterior sampling of the stiffness scaling parameter $\boldsymbol{\theta}_d$ for the monitoring stage.

We know from Section 3.3 that if the Markov chain created by the GS algorithms is ergodic, samples from the marginal distributions $p(\boldsymbol{\theta}_u|\boldsymbol{\mathcal{D}}_u)$ and $p(\boldsymbol{\theta}_d|\boldsymbol{\mathcal{D}}_d)$ are readily obtained by simply examining the GS samples $\boldsymbol{\theta}_u^{(n)}$ and $\boldsymbol{\theta}_d^{(n)}$ for larger $n$ beyond the burn-in period. Using samples from the marginal posterior PDF $p(\boldsymbol{\theta}_u|\boldsymbol{\mathcal{D}}_u)$ at the calibration stage, we are able to effectively take the uncertainty of $\boldsymbol{\theta}_u$ into account and draw samples from the posterior PDF $p(\boldsymbol{\theta}_d|\hat{\boldsymbol{\psi}}_d, \hat{\boldsymbol{\omega}}_d^2, \hat{\boldsymbol{\psi}}_u, \hat{\boldsymbol{\omega}}_u^2)$ which is conditional on modal data from both the monitoring and calibration stages. The PDF $p(\boldsymbol{\theta}_d|\hat{\boldsymbol{\psi}}_d, \hat{\boldsymbol{\omega}}_d^2, \hat{\boldsymbol{\psi}}_u, \hat{\boldsymbol{\omega}}_u^2)$ is expressed as:

$$p(\boldsymbol{\theta}_d|\hat{\boldsymbol{\psi}}_d, \hat{\boldsymbol{\omega}}_d^2, \hat{\boldsymbol{\psi}}_u, \hat{\boldsymbol{\omega}}_u^2) = \int p(\boldsymbol{\theta}_u, \boldsymbol{\theta}_d|\hat{\boldsymbol{\psi}}_d, \hat{\boldsymbol{\omega}}_d^2, \hat{\boldsymbol{\psi}}_u, \hat{\boldsymbol{\omega}}_u^2)\, d\boldsymbol{\theta}_u$$

$$= \int p(\boldsymbol{\theta}_d|\hat{\boldsymbol{\psi}}_d, \hat{\boldsymbol{\omega}}_d^2, \boldsymbol{\theta}_u) p(\boldsymbol{\theta}_u|\hat{\boldsymbol{\psi}}_u, \hat{\boldsymbol{\omega}}_u^2)\, d\boldsymbol{\theta}_u \tag{35}$$

It is known from the sampling theory that the extracted stiffness samples $\boldsymbol{\theta}_d^{(n)}$ from sample pairs $\left\{\boldsymbol{\theta}_u^{(n)}, \boldsymbol{\theta}_d^{(n)}\right\}, n = 1, \dots, N$, where the sample pairs are generated according to the joint posterior PDF $p(\boldsymbol{\theta}_u, \boldsymbol{\theta}_d|\hat{\boldsymbol{\psi}}_d, \hat{\boldsymbol{\omega}}_d^2, \hat{\boldsymbol{\psi}}_u, \hat{\boldsymbol{\omega}}_u^2)$, are distributed as the marginal posterior PDF $p(\boldsymbol{\theta}_d|\hat{\boldsymbol{\psi}}_d, \hat{\boldsymbol{\omega}}_d^2, \hat{\boldsymbol{\psi}}_u, \hat{\boldsymbol{\omega}}_u^2)$. For generating sample pairs $\left\{\boldsymbol{\theta}_u^{(n)}, \boldsymbol{\theta}_d^{(n)}\right\}, n = 1, \dots, N$, we first get $\left\{\boldsymbol{\theta}_u^{(n)}\right\}, n = 1, \dots, N$ by independently resampling from the Markov chain samples generated from $p(\boldsymbol{\theta}_u|\boldsymbol{\mathcal{D}}_u)$ by Gibbs sampling for the calibration stage, and then we collect the samples $\left\{\boldsymbol{\theta}_d^{(n)}\right\}, n = 1, \dots, N$, generated from the PDF $p(\boldsymbol{\theta}_d|\hat{\boldsymbol{\psi}}_d, \hat{\boldsymbol{\omega}}_d^2, \boldsymbol{\theta}_u)$ by Gibbs sampling for the monitoring stage, where we incorporate the $n^{th}$ sample $\boldsymbol{\theta}_u^{(n)}$ in the $n^{th}$ iteration sequentially. By implementing this procedure, the generation of posterior samples $\left\{\boldsymbol{\theta}_d^{(n)}\right\}, n = 1, \dots, N$, which characterize the structural damage information, is achieved.

Note that if we have large amount of identified modal datasets (i.e. $N_s \gg 1$), then Laplace's approximation can be effectively used in (35) to approximate that PDF $p(\boldsymbol{\theta}_d|\hat{\boldsymbol{\psi}}_d, \hat{\boldsymbol{\omega}}_d^2, \hat{\boldsymbol{\psi}}_u, \hat{\boldsymbol{\omega}}_u^2)$ by $p(\boldsymbol{\theta}_d|\tilde{\boldsymbol{\theta}}_u, \hat{\boldsymbol{\psi}}_d, \hat{\boldsymbol{\omega}}_d^2)$, where $\tilde{\boldsymbol{\theta}}_u$ is the MAP estimate of $p(\boldsymbol{\theta}_u|\hat{\boldsymbol{\psi}}_u, \hat{\boldsymbol{\omega}}_u^2)$ from the calibration stage. This assumption was made in [15].

### *4.2 Evaluation of damage probability*

The probabilistic evaluation of any possible damage, including its location and severity, is achieved by computing the probability that any stiffness parameter of a substructure has decreased more than a prescribed fraction $f$. Using the samples of $\boldsymbol{\theta}$ obtained from the monitoring stage (possibly damaged) and the calibration stage (undamaged), we estimate the probability of damage for the $jth$ substructure using the following approximation (see [30] and [31]):

$$P_j^{dam}(f) = P(\theta_{d,j} < (1-f)\theta_{u,j}|\hat{\boldsymbol{\psi}}_d, \hat{\boldsymbol{\omega}}_d^2, \hat{\boldsymbol{\psi}}_u, \hat{\boldsymbol{\omega}}_u^2)$$

$$\approx \frac{1}{N}\sum_{n=1}^{N} \mathbf{I}\left[\theta_{d,j}^{(n)} < (1-f)\theta_{u,j}^{(n)}\right] \tag{36}$$

where $\mathbf{I}[\cdot]$ is the indicator function, which is unity if the condition is satisfied, otherwise it is zero; $\left(\boldsymbol{\theta}_u^{(n)}, \boldsymbol{\theta}_d^{(n)}\right)$ denote the $n^{th}$ sample pairs of the stiffness scaling parameter for the calibration and monitoring stages, respectively, which are generated according to the joint posterior PDF $p\left(\boldsymbol{\theta}_u, \boldsymbol{\theta}_d | \widehat{\boldsymbol{\psi}}_d, \widehat{\boldsymbol{\omega}}_d^2, \widehat{\boldsymbol{\psi}}_u, \widehat{\boldsymbol{\omega}}_u^2\right)$ in (35).

## 5. COMPARISON OF NEW ALGORITHMS WITH THAT IN [15]

In [15], a fast sparse Bayesian learning algorithm was proposed for structural stiffness identification purposes, where the analytical solution of the posterior PDF of the stiffness scaling parameter $\boldsymbol{\theta}$ is computed. All uncertain parameters except $\boldsymbol{\theta}$ are collected in the vector $\boldsymbol{\delta} = [(\boldsymbol{\omega}^2)^T, \rho^T, \boldsymbol{\phi}^T, \eta, \boldsymbol{\alpha}^T, \beta, b_0]^T$ as 'nuisance' parameters, which are treated by using Laplace's approximation method (their posterior uncertainties are effectively ignored). For estimation of hyper-parameters in $\boldsymbol{\delta}$, we assume that the model class $\mathcal{M}(\boldsymbol{\delta})$ is globally identifiable based on the available data $\{\widehat{\boldsymbol{\omega}}^2, \widehat{\boldsymbol{\psi}}, \widehat{\boldsymbol{\theta}}_u\}$, meaning here that the likelihood $p(\widehat{\boldsymbol{\omega}}^2, \widehat{\boldsymbol{\psi}}, \widehat{\boldsymbol{\theta}}_u | \boldsymbol{\delta})$ has a unique global maximum over $\boldsymbol{\delta}$ and then so does the posterior at $\widetilde{\boldsymbol{\delta}}$ (the MAP value of $\boldsymbol{\delta}$). The excellent performance of the proposed method in [15] has been verified with synthetic data and the IASC–ASCE Phase II experimental brace benchmark data. However, there are several important issues that needed further exploration for reliable structural system identification and they have been explicitly addressed in the proposed Gibbs sampling algorithm. In the next two subsections, the theoretical benefits for real applications of the proposed Gibbs sampling algorithms are presented, followed by a comment on computational costs.

### *5.1 Theoretical benefits for real applications*

For structural identification of real systems where there are large modeling errors, it is challenging to confirm that there is a unique maximum with a sharp peak for the high-dimensional system modal parameter vectors $\{(\boldsymbol{\omega}^2)^T, \boldsymbol{\phi}^T\}$. In addition, the pseudo data $\widehat{\boldsymbol{\theta}}_u$ used in [15] is based on the assumption that it is a unique MAP estimate of $\boldsymbol{\theta}$ at the calibration stage due to a large amount of time-domain vibration data that allows multiple sets of identified modal parameters to be collected. Sometimes there may not be sufficient data available to get reliable updating results with a unique peak and small uncertainties for $\boldsymbol{\theta}$. Therefore, it is useful to explore the uncertainties in the identified system modal parameters and stiffness parameters from the calibration stage, as done in the proposed Gibbs sampling algorithms, which can provide a full characterization of the posterior uncertainty, even when the model class is not globally identifiable [22,23].

For the updating of the stiffness scaling parameters $\boldsymbol{\theta}$ and system modal parameters $\boldsymbol{\omega}^2$ and $\boldsymbol{\phi}$, the corresponding model classes $\mathcal{M}(\boldsymbol{\alpha}, \beta), \mathcal{M}(\rho, \beta)$ and $\mathcal{M}(\eta, \beta)$ (for Algorithm 1), and $\mathcal{M}(\boldsymbol{\gamma}, b_0), \mathcal{M}(\upsilon, b_0)$ and $\mathcal{M}(\tau, b_0)$ (for Algorithm 2) are investigated. When learning the corresponding hyper-parameters $\boldsymbol{\alpha}, \rho$ and $\eta$ by maximizing their posterior distribution (i.e., (A7) and (B6) in the Appendices), the application of Bayes' Theorem at the model class level involves a trade-off between the average data-fit of the model class and the information it extracts from the associated data. In other words, it automatically penalizes both models of $\boldsymbol{\theta}$ ($\boldsymbol{\omega}^2$or $\boldsymbol{\phi}$) that "under-fit" or "over-fit" the associated data $\widehat{\boldsymbol{\theta}}_u$ ($\widehat{\boldsymbol{\omega}}^2$or $\widehat{\boldsymbol{\psi}}$), therefore obtaining reliable updating results for the three parameter vectors, which is the Bayesian Ockham Razor [1] at work.

It was found that the algorithms in [15] suffer from a robustness problem where the system identification performance is sensitive to the selection of the equation error precision $\beta$. In Algorithm 2, by marginalizing over $\beta$ directly to remove it from the posterior distributions, we get the Student-t conditional PDFs that can be sampled in each step of the Gibbs sampling. The Student-t PDFs have heavier tails than the Gaussian PDFs sampled in Algorithm 1 and so the algorithm is robust to noise and outliers. The numerical and experimental results given later support this conclusion because Algorithm 2 outperforms all other algorithms for comparison.

*5.2 Comparison of the computational costs*

The computation is much more intensive for the proposed Gibbs sampling algorithms than the fast algorithm in [15], which is a sacrifice for gaining full posterior uncertainty quantification. Therefore, the choice of which method to use in system identification applications is a trade-off between the accuracy of the uncertainty quantification and the computation time. A comparison of the computation cost of the two Gibbs Sampling algorithms is given later in the illustrative examples. It shows that Algorithm 2 requires less iterations to make the parameter distributions well characterized since the number of effective dimensions for the Gibbs samplers is reduced from four in Algorithm 1 to three in Algorithm 2.

## 6. ILLUSTRATIVE EXAMPLES

The performance of the proposed GS algorithms is illustrated with the IASC-ASCE Phase II SHM benchmarks, first applying them to the same unbraced damage cases with simulated data as in [30] for comparison purposes. Then the brace damages cases from both the simulated and experimental data are used to verify the applicability of the GS algorithms in distinct cases with small and large modeling errors, respectively.

### *6.1 Example I: Unbraced damage cases in simulated Phase II benchmark problem*

In the first example, the proposed GS algorithms are applied to the beam-column joint damage cases DP1U and DP2U in the simulated unbraced Phase II benchmark structure [41], and used for comparing with the method in [30], who focus on the same damage patterns in their Example 2. Figure 2 depicts a diagram of the analytical benchmark model that shows its dimensions, in which the *x*-direction is the strong direction of the columns, and the locations of the connection damage. It is a 3D four-story, two-bay by two-bay steel unbraced structural model with 120 DOF. The reader is referred [33] for detailed information, including the nominal properties of the structural elements, damage patterns, and so on. The simulated ambient-vibration data are generated by running a Matlab program of the finite-element model subjected to an excitation of broadband stationary wind forces at each floor. The measurements are available at the center of each side of each floor with the directions parallel to either the positive *x*- or *y*-direction. Zero-mean Gaussian noise is also added with variance at a high level of 10% of the mean square of the synthetic measurements. The extraction of the modal parameters was achieved by using the modal identification procedure called MODE-ID [42,43] and the results were presented in [33]. In each experiment, the measured time series are divided into ten time segments of equal duration of 20s (10,000 sampling points with sampling frequencies of 500 Hz) and eight modes, consisting of the four translation modes in the $x$ direction and four translation modes in the y direction, are identified for each time segment.

For damage identification, we project the 120-DOF unbraced structural model onto a shear-building model with 36-DOF that assumes rigid floors in the $x$-$y$ plane and allows rotation along the $x$- and $y$-axes. For each floor, rotation along the $z$-axis and translations parallel to the $x$- and $y$-axes give three of the nine DOF. The remaining six DOF are given based on the assumption that nodes with the same $x$-coordinates or same $y$-coordinates have the same amount of rotation along the $y$-axis and $x$-axis, respectively. There are three stiffness parameters assigned for each story, corresponding to two rotational and one column stiffness scaling parameters, which give 12 parameters in total:

$$\mathbf{K}(\boldsymbol{\theta}) = \sum_s \theta_{cs}\overline{\mathbf{K}}_{cs} + \sum_s \sum_u \theta_{su}\overline{\mathbf{K}}_{su} \tag{37}$$

where $s = 1, \ldots, 4$ refers to the story number and $u = x, y$ indicates the axis along which the rotational stiffness is active. In (37), $\overline{\mathbf{K}}_{cs}$ and $\overline{\mathbf{K}}_{cs}$ are the "nominal" column and rotational stiffness matrices, respectively, so that the nominal value of each stiffness scaling parameter is 1.0 and these matrices are computed based on the model assumptions for the original undamaged structure. Based on this modeling, the true stiffness ratio values for $\theta_{1,y}$ and $\theta_{2,y}$ in the DP1U damage case are 50% and 66.7% of the undamaged values, respectively, while the true ratio value for $\theta_{1,y}$ in the DP2U case is 66.7% of the undamaged value.

By running Algorithms 1 and 2 in the calibration (undamaged) and monitoring (possibly damaged) stages, the GS Markov chain samples are generated for the model parameters of interest. Inspired by [30], we examine the practical ergodicity of the Markov chain by conducting five parallel GSs to obtain five independent Markov chains. All of the five Markov Chains converged to the same region of the parameter space, and so we assume that the GS Markov chain generated is ergodic. We first present the Markov chain samples of the eight rotational stiffness scaling parameters for the undamaged case and DP2U damage case in Figures 3 and 4, respectively. For both cases, it is clearly seen that the Markov chain reaches its stationary state after a burn-in period of roughly 6000 samples for Algorithm 1, while only 2000 samples are required for Algorithm 2. The higher efficiency of Algorithm 2 seems to be reasonable since the number of effective dimensions for Gibbs samplers is reduced from four to three by integrating out the equation error precision parameter $\beta$ in Algorithm 2. In fact, it is known that the performance of Bayesian learning is sensitive to the selection of the equation error precision $\beta$ [15] and so the robustness for parameter inference should be enhanced significantly if we marginalize over this parameter directly to remove it from the posterior distributions as done in Algorithm 2. Indeed, Figures 3 and 4 show that this algorithm give correct results for the undamaged (scaling≈1.0) and damaged substructure (scaling≈0.6 for $\theta_{1,y}$ in the DP2U).

Figure 5 shows the probabilities that the eight rotation stiffness parameters in the two damage cases have been reduced by a certain damage fraction $f$ compared to the undamaged state based on Algorithms 1 and 2. These probabilities are estimated using (36) based on different numbers of the Markov chain samples (1000, 2000, 3000 and 4000) after burn-in period (6000 samples). It is observed that the damage probabilities estimated from different numbers of samples are almost the same for Algorithm 2, while there is an obvious difference in the results for Algorithm 1, demonstrating the GS Markov chain samples created by Algorithm 2 represent the target PDF better. For the damage detection performance, all the actual damage locations (corresponding to stiffness scaling parameters $\theta_{1,y}$ and $\theta_{2,y}$ for the DP1U case and $\theta_{1,y}$ for the DP2U case) are correctly localized for both algorithms, but for the DP1U case there is a false damage detection in the substructure corresponding to stiffness scaling parameter $\theta_{1,x}$ for Algorithm 1. The inferred damage extents are also reliable for Algorithm 2: the posterior medians of the estimated damage corresponding to a damage probability of 0.5 is about 50% loss in $\theta_{1,y}$ and $\theta_{2,y}$ for DP1U and 30% loss in $\theta_{1,y}$ for DP2U, which are close to their actual values. While for the undamaged substructures, the median losses of rotational stiffness are around zero.

We now compare Figures 3-5 to the results from the method in [30]. Comparing the results for the Markov chain samples, the determination of the burn-in period can be made with more confidence due to the fact that the spread of the post burn-in samples is smaller for our new GS methods. This is presumably a benefit from the hierarchical Bayesian model employed in the formulation, which allows Bayesian learning of the rate parameter $b_0$ for the prior PDF of equation error precision $\beta$, as in (25), (27e), (30e) and (33e), rather than simply setting its value to create a Jeffreys' non-informative prior ($a_0 = 0$ and $b_0 = 0$ are assigned for the Gamma prior of $\beta$ in [30]). The learning of $b_0$ suppresses both the prior and posterior uncertainties of $\beta$, and so the posterior uncertainties for the model parameters $\{\boldsymbol{\phi}, \boldsymbol{\omega}^2, \boldsymbol{\theta}\}$. This learning of the hyperparameters in the formulation also explains why the undamaged substructures are identified with much higher confidence in our new proposed methods than the method in [30] when comparing the damage probability curves. The sparse models of $\Delta\boldsymbol{\theta} = \boldsymbol{\theta} - \boldsymbol{\theta}_u$ are produced by learning the hyper-parameter $\boldsymbol{\alpha}$ in (21) (or $\boldsymbol{\gamma}$ in (34)) in the sparse Bayesian learning formulation [15,17], where the MAP value $\tilde{\alpha}_j \to 0$ $\left(\text{or } \tilde{\gamma}_j \to 0\right)$ implies that $(\boldsymbol{\Sigma}_{\boldsymbol{\theta}})_{jj} \to 0$ (or $(\boldsymbol{\Lambda}_{\boldsymbol{\theta}})_{jj} \to 0$ ) and so the GS sample $\boldsymbol{\theta}_j \to (\boldsymbol{\theta}_u)_j$.

### *6.2 Example II: Braced damage case in simulated and experimental Phase II benchmark problems*

In the second example, we apply the proposed GS algorithms to the braced damage cases in the IASC-ASCE Phase II Benchmarks, first using the simulated data [41], and then using the experimental data [44]. The benchmark structural model here refers to a four-story, two-bay by two-bay steel braced-frame, which is depicted diagrammatically in Figure 6, along with its dimensions. A detailed description of the benchmark structure in the simulated and experimental studies can be found in [33].

For the *simulated* data case, four brace damage patterns are considered: DP1B, DP2B, DP3B and DP3Bu, which are simulated by reducing the elastic moduli of certain braces. RB is the undamaged structure which serves to provide the calibration stage tests in our theory. In each of these five cases, simulated time-domain data are generated from the structural model, which is subjected to broadband ambient-vibration excitations at each floor, and 10% simulated zero-mean Gaussian noise is added at the measured DOF. Only the results of the partial-sensor scenario are presented in the example, where measurements are available only at the third floor at the nodes 20,22,24,26 in Figure 6 and the roof at the nodes 38,40,42,44. Ten sets of independent estimates of the modal data, including four modes in the strong ($x$) direction and four modes in the weak ($y$) direction in each set, are extracted in [33] and are utilized in this study.

The *experimental* benchmark study includes five brace damage configurations: Config. 2, Config. 3, Config. 4, Config. 5 and Config. 6, by removal of certain braces of the structure. Config. 1 is the undamaged structure which serves to provide the calibration stage. For each configuration, acceleration data is obtained on the experimental structure by impact of a sledgehammer, where sensors are installed at the center at each floor sensing the accelerations in the $+y$ direction and the $+y$ and $-y$ faces of all floors sensing the accelerations in the $+x$ direction. For modal identification, a total of three sets of independent estimates of the experimental modal parameters are extracted using MODE-ID [42,43] and five modes ($N_m = 5$), consisting of the first and second translation modes in the $x$ and $y$ directions and the first torsion mode, are identified from each time segment. The reader is referred to [33] for detailed information.

For locating the faces sustaining brace damage, a 3-D 12-DOF shear-building model that assumes rigid floors is employed, where the three DOF for each floor are the translations parallel to the $x$ and $y$ axes and rotation about the $z$ axis vertically through the center of the structure. The stiffness matrix **K** is then parameterized as:

$$\mathbf{K}(\boldsymbol{\theta}) = \mathbf{K}_0 + \sum_s \sum_f \theta_{sf} \overline{\mathbf{K}}_{sf} \quad (38)$$

where $s$=1,…,4 refers to the story number and $f$= $'+x', '-x', '+y', '-y'$ indicates the faces of the respective floor. Four stiffness parameters are used for each story to give a stiffness scaling parameter vector $\boldsymbol{\theta}$ with 16 components. The "nominal" stiffness matrices $\overline{\mathrm{K}}_{\mathrm{sf}}$ are defined to make the nominal value of each $\theta_{sf}$ to be 1.0. Based on this modeling, the sub-structure stiffness reductions for the four brace damage patterns in the simulated benchmark are summarized as follows: 1) DP1B: 11.3% reduction in $\theta_{1,+y}$ and $\theta_{1,-y}$; 2) DP2B: 5.7% reduction in $\theta_{1,+y}$ and $\theta_{1,-y}$; 3) DP3B: 11.3% reduction in $\theta_{1,+y}$ and $\theta_{1,-y}$; 5.7% reduction in $\theta_{3,+y}$ and $\theta_{3,-y}$; 4) DP3Bu: 11.3% reduction in $\theta_{1,-y}$ and 5.7% reduction in $\theta_{3,-y}$. For the experimental benchmark, the stiffness reduction ratio values for $\theta_{1,-y}$, $\theta_{2,-y}$, $\theta_{3,-y}$ and $\theta_{4,-y}$ for Config. 2 and $\theta_{1,-y}$ for Config. 6 are 45.1% of the undamaged values while the stiffness reduction ratio values for $\theta_{1,-y}$, $\theta_{2,-y}$, $\theta_{3,-y}$ and $\theta_{4,-y}$ for Config. 3, $\theta_{1,-y}$ and $\theta_{4,-y}$ for Config. 4, and $\theta_{1,-y}$ for Config. 5 are all 22.6%.

We first apply the GS algorithms in the calibration (undamaged) stage. All 16 components $\theta_j$ of $\boldsymbol{\theta}$ are updated where $\theta_1$,…, $\theta_{16}$ corresponds to the order of $\theta_{1,+x}$, ..., $\theta_{4,+x}$, $\theta_{1,+y}$, ..., $\theta_{4,+y}$, $\theta_{1,-x}$, ..., $\theta_{4,-x}$, $\theta_{1,-y}$, ..., $\theta_{4,-y}$. The Markov chain samples of the eight stiffness scaling parameters at the $-x$ face and $-y$ face in each story are plotted in Figure 7 for the undamaged case in the simulated benchmark problem. Note that the y-scale differs from Figures 3 and 4 since the posterior sampling variance is smaller in Figure 7. Similar to Figures 3 and 4, Algorithm 2 demonstrates that a much smaller number of Markov Chain samples is required (roughly 200 samples) for reaching the stationary state while Algorithm 1 takes much longer (roughly 4000 samples). On the other hand, for the experimental benchmark, we found that the samples of $\boldsymbol{\theta}$ spread out in an unacceptably wide area of the parameter space if the number of components for parameter $\boldsymbol{\theta}$ is selected to be 16 for the undamaged case, showing that the available information contained in the three sets of identified modal data is poor to support a reliable inference of the stiffness scaling parameters $\boldsymbol{\theta}$. We then assume that the stiffness scaling parameters of $+y$ and $-y$ faces for all floors, and $+x$ and $-x$ faces for all floors, are identical, in order to reduce the number of stiffness model parameters to two. The results of the Markov chain for the two stiffness model parameters are presented in Figure 8, which shows that the spread becomes reasonable for the group of samples. However, it is observed that the target PDF characterized by the GS samples has large uncertainties and very few samples are required to reach the stationary state, presumably because the large uncertainties in the model parameters conceal the fluctuation of the samples in the possibly transient period. We only collect the Markov chain samples $\boldsymbol{\theta}_u$ when the time step is larger than 4000, which are used as pseudo-data for $\boldsymbol{\theta}$ in the monitoring stage.

We next focus on the results in the monitoring stage. In Figures 9 and 10, all the samples generated from Algorithms 1 and 2, excluding those in the burn-in period (4000 samples), are plotted in the $\{\theta_{1,-y}, \theta_{2,-y}\}$ and $\{\theta_{3,-y}, \theta_{4,-y}\}$ spaces for the DP3B case (simulated benchmark) and Config. 5 case (experimental benchmark), respectively. For both algorithms applied to the simulated data, it is clearly seen from Figure 9 that the reduction of the sub-structure stiffnesses at the $-y$ face in the first and third stories are correctly detected for DP3B case. Moreover, the amount of the identified stiffness loss is approximately correct, i.e., using the mean of the sample values, there is about 12% loss in $\theta_{1,-y}$, and 5% loss in $\theta_{3,-y}$. Similarly, Figure 10 shows that the stiffness reduction in the Config. 5 case is also corrected identified and quantified as far as the sample means are concerned. However, the posterior sample variance is large because the samples are spread out much more for the Config. 5 case as expected, due to large modeling errors for the real data case.

Given the Markov chain samples in the calibration and monitoring stages, we use (36) to evaluate the probabilities of damage for each substructure and the results for the four damage cases in the simulated benchmark are shown in Figure 11. All the brace damage substructures are reliably detected in both qualitative and quantitative ways for both methods. For the undamaged substructures, the posterior medians of the estimated damage corresponding to a damage probability of 0.5 are around zero, although in some cases some undamaged substructures show an increase in stiffness with large probabilities that is unrealistic. No occurrence of false damage detection is observed for both methods. This is an advantage of the proposed sparse Bayesian formulation which can better quantify the uncertainty of the unchanged components of $\boldsymbol{\theta}$.

For the experimental benchmark, the damage probability curves corresponding to the sixteen stiffness scaling parameters $\boldsymbol{\theta}$ are plotted in Figure 12, for Configs. 2–6. To demonstrate the robustness enhancement of our proposed GS algorithms where the full uncertainties in the model parameter of interest are treated, the results are compared with our previous method [15], where the damage probability curves are evaluated using a Gaussian approximation (the number of stiffness model parameters is also selected to be two in the calibration stage as in this work). Although more undamaged substructures are observed having a significant stiffness increase than in the simulated data case, all actual damaged substructures are clearly shown to have a large damage probability when their stiffness scaling parameters have been reduced by the actual fraction for all of the three methods, except for Config 6. In Config 6, the posterior medians of the estimated damage for the damaged substructure $\theta_{2,+x}$ are underestimated by approximately 20% for all three methods, compared to the real stiffness reduction ratio of 45.1%. With respect to the performance comparison among the three methods, the occurrence of false damage detections seems to be more unlikely for Algorithm 2 than the other two, due to the smaller uncertainty in the model parameters. For example, there are several undamaged substructures having comparable damage probabilities with those of the damaged substructures in Configs. 3 and 6 for the method in [15] and in Configs. 5 and 6 for Algorithm 1, even for large damage fractions. While for Algorithm 2, the possible false damage detection occurs in the substructure corresponding to $\theta_{4,-x}$ in Config. 5, but its damage probability is much smaller than those of two actually damaged substructures $\theta_{1,-y}$ and $\theta_{4,-y}$ where the stiffness reduction is larger than 40%.

## 7. CONCLUSION AND FUTURE WORK

We have presented and applied two Gibbs sampling algorithms for Bayesian system identification based on incomplete modal data and where the spatial distribution of structural stiffness change from its calibration value is sparse. The adoption of Gibbs sampling is motivated by its important advantage that the effective dimension of the Gibbs samplers only depends on the small number of parameter groups for drawing samples, and so the algorithmic efficiency does not degrade with larger dimensions of the uncertain parameter space. The algorithms are based on a similar hierarchical sparse Bayesian model that we have used in previous research and that allows all sources of uncertainty and correlation to be learned from the data to produce more reliable system identification results.

The two proposed Gibbs sampling algorithms differ in their strategies to deal with the posterior uncertainty of the equation-error precision parameter: Algorithm 1 samples the posterior PDF for the equation error precision parameter $\beta$ directly whereas Algorithm 2 marginalizes it out to remove it as a "nuisance" parameter. When applied to simulated data where there are smaller modeling errors, it is found that the Markov chain reaches its stationary state much faster for Algorithm 2. Also, for the experimental benchmark problem where there are larger modeling errors, the spread for the group of GS samples generated from Algorithm 2 is more concentrated. The performance of Algorithm 2 for the challenging IASC-ASCE Phase II experimental benchmark is better than that of Algorithm 1 and our previous proposed sparse Bayesian learning method in [15].

With regard to the comparison between our proposed Gibbs Sampling algorithms and that in [30], several appealing features have been demonstrated by the new algorithms in terms of theoretical and numerical aspects. First, system frequencies are introduced as parameters to be identified in order to represent the actual natural frequencies of the structural system along with the system mode shapes, and the eigenvalue equations of the structural model are used only in the prior probability distribution to provide "soft" constraints. Second, all hyper-parameters are learned solely from the data available and hence there is no parameter tuning required. Third, model sparseness in the inferred stiffness losses is produced by the sparse Bayesian learning framework, which is useful for model regularization to alleviate the ill-posedness in the structural system identification and damage assessment. Fourth, the equation-error precision parameter is marginalized over analytically in the formulation for Algorithm 2, which reduces the effective dimension for the Gibbs sampler, which has been shown to be advantageous for reducing the posterior sample variances. Fifth, the stiffness identification results are totally scale-invariant in the new algorithms, i.e., they are independent of the unit selections for the mass and stiffness matrices, as well as the scaling of the measured mode shapes.

For future studies, a potentially useful avenue could be to utilize the proposed Bayesian system identification framework in a sequential manner to track the temporal behavior of the structural model parameters, where the model updating results from the previous time can be effectively incorporated for updating at the current time using the sparse Bayesian learning formulation.

**Nomenclature**

$N_m$ = Number of extracted modes in the modal identification we use $(i = 1, \dots, N_m)$

$N_o$ = Number of measured degrees of freedom

$N_d$ = Number of degrees of freedom of the identification model we use $(k = 1, \dots, N_d)$

$N_s$ = Number of time segments of measured modal data we use $(r = 1, \dots, N_s)$

$N_\theta$ = Number of substructures considered we use $(j = 1, \dots, N_\theta)$

$\mathbf{M}, \mathbf{K}$ = Mass and stiffness matrices of the structural model for identification

$\boldsymbol{\theta}$ = Structural stiffness scaling parameters

$\boldsymbol{\theta}_u$ = Structural stiffness scaling parameters for undamaged calibration stages

$\boldsymbol{\phi}_r, \boldsymbol{\omega}_r^2$ = System mode shape and system natural frequencies of the $rth$ mode

$\beta$ = Equation-error precision parameter

$\hat{\boldsymbol{\omega}}_{r,i}^2, \hat{\boldsymbol{\psi}}_{r,i}$ = MAP estimates of modal frequency and mode shape of $i^{th}$ mode from the $r^{th}$ data segment from modal identification

$\boldsymbol{\Gamma}$ = Matrix that picks the measured degrees of freedom from the system mode shape

$\mathbf{T}$ = Matrix relating the vector of $N_s$ sets of identified natural frequencies $\hat{\boldsymbol{\omega}}^2$ and the system natural frequencies $\boldsymbol{\omega}^2$

$\boldsymbol{\alpha}$ = Variance parameter for the likelihood function of structural stiffness scaling parameters $\boldsymbol{\theta}$

$\eta, \rho$ = Measurement-error variance parameters for system mode shapes and natural frequencies

**Appendix A: Derivation of conditional posterior PDFs for Algorithm 1**

We derive the conditional posterior PDFs $p(\boldsymbol{\phi}|\mathbf{y},\boldsymbol{\omega}^2,\boldsymbol{\theta},\beta)$, $p(\boldsymbol{\omega}^2|\mathbf{y},\boldsymbol{\phi},\boldsymbol{\theta},\beta)$ and $p(\boldsymbol{\theta}|\mathbf{y},\boldsymbol{\phi},\boldsymbol{\omega}^2,\beta)$ for Gibbs sampling in this Appendix.

We rewrite the conditional prior PDFs in (6a), (6b), (6c) and marginal prior PDFs in (5a), (5b), (5c) in the following general form:

$$p(\mathbf{x}_1|\mathbf{x}_2,\mathbf{x}_3,\beta) = \mathcal{N}(\mathbf{x}_1|(\mathbf{E}^T\mathbf{E})^{-1}\mathbf{E}^T\mathbf{r},(\beta\mathbf{E}^T\mathbf{E})^{-1}) \tag{A1}$$

$$p(\mathbf{x}_2,\mathbf{x}_3|\beta) = c(2\pi/\beta)^{(M-N_dN_m)/2}|\mathbf{E}^T\mathbf{E}|^{-1/2}\exp\left\{-\frac{\beta}{2}(\mathbf{r}^T\mathbf{r}-\mathbf{r}^T\mathbf{E}\,(\mathbf{E}^T\mathbf{E})^{-1}\mathbf{E}^T\mathbf{r})\right\} \tag{A2}$$

where $\mathbf{E}$ and $\mathbf{r}$ are a fixed matrix and vector which depend on the other parameters $\mathbf{x}_2$ and $\mathbf{x}_3$; $\beta$ is the eigenequation-error precision parameter and $M$ is the dimensionality of $\mathbf{x}_1$. Each choice of $\mathbf{x}_1$, $\mathbf{x}_2$ and $\mathbf{x}_3$ is a permutation of $\boldsymbol{\phi},\boldsymbol{\omega}^2$ and $\boldsymbol{\theta}$.

Similarly, the likelihood functions in (9), (10) and (11) can be written in the following general form:

$$p(\mathbf{y}|\mathbf{x}_1,\boldsymbol{\kappa}) = \mathcal{N}(\mathbf{y}|\boldsymbol{\Theta}\mathbf{x}_1,\mathbf{L}) \tag{A3}$$

where vector $\mathbf{y}$ is the available data, $\boldsymbol{\Theta}$ is a matrix and $\mathbf{y}=\boldsymbol{\Theta}\mathbf{x}_1+\boldsymbol{\varepsilon}$, where the Gaussian prediction error $\boldsymbol{\varepsilon}$ has covariance matrix $\mathbf{L}(\boldsymbol{\kappa})$, a function of uncertain hyper-parameter $\boldsymbol{\kappa}$, which is $\boldsymbol{\alpha},\rho$ and $\eta$ in Subsection 3.1.

The conditional posterior PDF over parameter vector $\mathbf{x}_1$ is obtained by using Laplace's asymptotic approximation [20] to integrate the "nuisance" hyper-parameters $\boldsymbol{\kappa}$ out:

$$p(\mathbf{x}_1|\mathbf{y},\mathbf{x}_2,\mathbf{x}_3,\beta) = \int p(\mathbf{x}_1|\mathbf{y},\mathbf{x}_2,\mathbf{x}_3,\beta,\boldsymbol{\kappa})\,p(\boldsymbol{\kappa}|\mathbf{y},\mathbf{x}_2,\mathbf{x}_3,\beta)d\boldsymbol{\kappa} \approx p(\mathbf{x}_1|\mathbf{y},\mathbf{x}_2,\mathbf{x}_3,\beta,\tilde{\boldsymbol{\kappa}}) \tag{A4}$$

where $\tilde{\boldsymbol{\kappa}} = \arg\max p(\boldsymbol{\kappa}|\mathbf{y},\mathbf{x}_2,\mathbf{x}_3,\beta)$. The above approximation is based on the assumption that the posterior $p(\boldsymbol{\kappa}|\mathbf{y},\mathbf{x}_2,\mathbf{x}_3,\beta)$ has a unique maximum at $\tilde{\boldsymbol{\kappa}}$ (the MAP value of $\boldsymbol{\kappa}$).

The posterior PDF of $\mathbf{x}_1$ given the specified $\tilde{\boldsymbol{\kappa}}$ is computed by Bayes' theorem:

$$p(\mathbf{x}_1|\mathbf{y},\mathbf{x}_2,\mathbf{x}_3,\beta,\tilde{\boldsymbol{\kappa}}) \propto p(\mathbf{y}|\mathbf{x}_1,\tilde{\boldsymbol{\kappa}})p(\mathbf{x}_1|\mathbf{x}_2,\mathbf{x}_3,\beta) \propto \mathcal{N}(\mathbf{x}_1|\boldsymbol{\mu},\boldsymbol{\Sigma}) \tag{A5}$$

where we have combined the Gaussian prior $p(\mathbf{x}_1|\mathbf{x}_2,\mathbf{x}_3,\beta)$ in (A1) and the Gaussian likelihood $p(\mathbf{y}|\mathbf{x}_1,\tilde{\boldsymbol{\kappa}})$ in (A3) to get the Gaussian posterior PDF $\mathcal{N}(\mathbf{x}_1|\boldsymbol{\mu},\boldsymbol{\Sigma})$ with mean and covariance matrix:

$$\boldsymbol{\mu}=\boldsymbol{\Sigma}(\boldsymbol{\Theta}^T\tilde{\mathbf{L}}^{-1}\mathbf{y}+\beta\mathbf{E}^T\mathbf{r}) \tag{A6a}$$

$$\boldsymbol{\Sigma}=(\beta\mathbf{E}^T\mathbf{E}+\boldsymbol{\Theta}^T\tilde{\mathbf{L}}^{-1}\boldsymbol{\Theta})^{-1} \tag{A6b}$$

where $\tilde{\mathbf{L}} = \mathbf{L}(\tilde{\boldsymbol{\kappa}})$.

This leaves the task of maximizing the PDF $p(\boldsymbol{\kappa}|\mathbf{y},\mathbf{x}_2,\mathbf{x}_3,\beta)$ to find the MAP value $\tilde{\boldsymbol{\kappa}}$. Using Bayes' Theorem and taking $\boldsymbol{\kappa}$ independent a priori of $\mathbf{x}_2,\mathbf{x}_3$ and $\beta$:

$$p(\boldsymbol{\kappa}|\mathbf{y},\mathbf{x}_2,\mathbf{x}_3,\beta) \propto p(\mathbf{y}|\mathbf{x}_2,\mathbf{x}_3,\beta,\boldsymbol{\kappa})\, p(\boldsymbol{\kappa})$$

$$\propto \int p(\mathbf{y}|\mathbf{x}_1,\boldsymbol{\kappa})\, p(\mathbf{x}_1|\mathbf{x}_2,\mathbf{x}_3,\beta) d\mathbf{x}_1 \cdot p(\boldsymbol{\kappa})$$

$$\propto \mathcal{N}(\mathbf{y}|\boldsymbol{\Theta}(\mathbf{E}^T\mathbf{E})^{-1}\mathbf{E}^T\mathbf{r}, \mathbf{L}+\beta^{-1}\boldsymbol{\Theta}(\mathbf{E}^T\mathbf{E})^{-1}\boldsymbol{\Theta}^T)\cdot p(\boldsymbol{\kappa}) \tag{A7}$$

The MAP estimates of $\boldsymbol{\kappa}$ can be obtained by maximizing the logarithm function of the PDF $p(\boldsymbol{\kappa}|\mathbf{y},\mathbf{x}_2,\mathbf{x}_3,\beta)$ without including the constants that do not depend on $\boldsymbol{\kappa}$. The objective function $\mathcal{L}(\boldsymbol{\kappa})$ is written as the following form:

$$\mathcal{L}(\boldsymbol{\kappa}) = -\frac{1}{2}\log|\mathbf{L}+\beta^{-1}\boldsymbol{\Theta}(\mathbf{E}^T\mathbf{E})^{-1}\boldsymbol{\Theta}^T| - \frac{1}{2}(\mathbf{y}-\boldsymbol{\Theta}(\mathbf{E}^T\mathbf{E})^{-1}\mathbf{E}^T\mathbf{r})^T(\mathbf{L}+\beta^{-1}\boldsymbol{\Theta}(\mathbf{E}^T\mathbf{E})^{-1}\boldsymbol{\Theta}^T)^{-1}(\mathbf{y}-\boldsymbol{\Theta}(\mathbf{E}^T\mathbf{E})^{-1}\mathbf{E}^T\mathbf{r}) \tag{A8}$$

First, following the Appendix in [15], we obtain the following equations using the determinant identity and Woodbury matrix identity, respectively:

$$\log|\mathbf{L}+\beta^{-1}\boldsymbol{\Theta}(\mathbf{E}^T\mathbf{E})^{-1}\boldsymbol{\Theta}^T| = \log(|\mathbf{L}||(\beta\mathbf{E}^T\mathbf{E})^{-1}||\beta\mathbf{E}^T\mathbf{E}+\boldsymbol{\Theta}^T\mathbf{L}^{-1}\boldsymbol{\Theta}|) = \log|\mathbf{L}| - \log|\beta\mathbf{E}^T\mathbf{E}| + \log|\boldsymbol{\Sigma}^{-1}| \tag{A9}$$

$$(\mathbf{y}-\boldsymbol{\Theta}(\mathbf{E}^T\mathbf{E})^{-1}\mathbf{E}^T\mathbf{r})^T(\mathbf{L}+\beta^{-1}\boldsymbol{\Theta}(\mathbf{E}^T\mathbf{E})^{-1}\boldsymbol{\Theta}^T)^{-1}(\mathbf{y}-\boldsymbol{\Theta}(\mathbf{E}^T\mathbf{E})^{-1}\mathbf{E}^T\mathbf{r})$$

$$= (\mathbf{y}-\boldsymbol{\Theta}(\mathbf{E}^T\mathbf{E})^{-1}\mathbf{E}^T\mathbf{r})^T(\mathbf{L}^{-1}-\mathbf{L}^{-1}\boldsymbol{\Theta}\boldsymbol{\Sigma}\boldsymbol{\Theta}^T(\mathbf{L}^{-1})^T)(\mathbf{y}-\boldsymbol{\Theta}(\mathbf{E}^T\mathbf{E})^{-1}\mathbf{E}^T\mathbf{r})$$

$$= (\mathbf{y}-\boldsymbol{\Theta}\boldsymbol{\mu})^T\mathbf{L}^{-1}(\mathbf{y}-\boldsymbol{\Theta}\boldsymbol{\mu}) + \beta(\boldsymbol{\mu}-(\mathbf{E}^T\mathbf{E})^{-1}\mathbf{E}^T\mathbf{r})^T\mathbf{E}^T\mathbf{E}(\boldsymbol{\mu}-(\mathbf{E}^T\mathbf{E})^{-1}\mathbf{E}^T\mathbf{r}) \tag{A10}$$

In general, the covariance matrix $\mathbf{L}(\boldsymbol{\kappa})$ is defined as a diagonal matrix. If the derivative of matrix $\mathbf{L}(\boldsymbol{\kappa})$ with respect to component $\kappa_m$ in $\boldsymbol{\kappa}$ is $\frac{\partial \mathbf{L}(\boldsymbol{\kappa})}{\partial \kappa_m} = \mathbf{X}_m$, then we have $\mathbf{L}^{-1}\mathbf{X}_m = \kappa_m^{-1}\mathbf{X}_m$ and $\mathbf{L}^{-2}\mathbf{X}_m = \kappa_m^{-2}\mathbf{X}_m$. Using (A9) and (A10), the derivative of $\mathcal{L}(\boldsymbol{\kappa})$ in (A8) with respect to $\kappa_m$ is given by:

$$\frac{\partial \mathcal{L}(\boldsymbol{\kappa})}{\partial \kappa_m}$$

$$= -\frac{1}{2}\frac{\partial}{\partial \kappa_m}[\log|\mathbf{L} + \beta^{-1}\boldsymbol{\Theta}(\mathbf{E}^T\mathbf{E})^{-1}\boldsymbol{\Theta}^T| + (\mathbf{y} - \boldsymbol{\Theta}(\mathbf{E}^T\mathbf{E})^{-1}\mathbf{E}^T\mathbf{r})^T(\mathbf{L} + \beta^{-1}\boldsymbol{\Theta}(\mathbf{E}^T\mathbf{E})^{-1}\boldsymbol{\Theta}^T)^{-1}(\mathbf{y} - \boldsymbol{\Theta}(\mathbf{E}^T\mathbf{E})^{-1}\mathbf{E}^T\mathbf{r})]$$

$$= -\frac{1}{2}\left[\frac{\partial \log|\mathbf{L}|}{\partial \kappa_m} + \frac{\partial \log\left|\boldsymbol{\Sigma}^{-1}\right|}{\partial \kappa_m} + (\mathbf{y} - \boldsymbol{\Theta}\boldsymbol{\mu})^T\frac{\partial \mathbf{L}^{-1}}{\partial \kappa_m}(\mathbf{y} - \boldsymbol{\Theta}\boldsymbol{\mu})\right]$$

$$= \mathrm{tr}(\mathbf{L}^{-1}\mathbf{X}_i) - \mathrm{tr}(\boldsymbol{\Sigma}\boldsymbol{\Theta}^T\mathbf{L}^{-2}\mathbf{X}_i\boldsymbol{\Theta}) - (\mathbf{y} - \boldsymbol{\Theta}\boldsymbol{\mu})^T\mathbf{L}^{-2}\mathbf{X}_m(\mathbf{y} - \boldsymbol{\Theta}\boldsymbol{\mu})$$

$$= \kappa_m^{-1}\mathrm{tr}(\mathbf{X}_i) - \kappa_m^{-2}\mathrm{tr}(\boldsymbol{\Sigma}\boldsymbol{\Theta}^T\mathbf{X}_i\boldsymbol{\Theta}) - \kappa_m^{-2}(\mathbf{y} - \boldsymbol{\Theta}\boldsymbol{\mu})^T\mathbf{X}_m(\mathbf{y} - \boldsymbol{\Theta}\boldsymbol{\mu}) \quad \text{(A11)}$$

Setting the derivative in (A11) to zero leads to the update formula for the MAP value $\tilde{\kappa}_m$:

$$\tilde{\kappa}_m = \frac{\mathrm{tr}(\boldsymbol{\Sigma}\boldsymbol{\Theta}^T\mathbf{X}_m\boldsymbol{\Theta}) + (\mathbf{y} - \boldsymbol{\Theta}\boldsymbol{\mu})^T\mathbf{X}_m(\mathbf{y} - \boldsymbol{\Theta}\boldsymbol{\mu})}{\mathrm{tr}(\mathbf{X}_m)} \quad \text{(A12)}$$

Note that when differentiating (A11) with respect to uncertain parameters $\boldsymbol{\kappa}$, the terms involving derivatives of $\boldsymbol{\mu}$ drop out since:

$$\beta(d\boldsymbol{\mu})^T\mathbf{E}^T\mathbf{E}(\boldsymbol{\mu} - (\mathbf{E}^T\mathbf{E})^{-1}\mathbf{E}^T\mathbf{r}) + \beta(\boldsymbol{\mu} - (\mathbf{E}^T\mathbf{E})^{-1}\mathbf{E}^T\mathbf{r})^T\mathbf{E}^T\mathbf{E}\cdot d\boldsymbol{\mu} + (-d\boldsymbol{\mu})^T\boldsymbol{\Theta}^T\mathbf{L}^{-1}(\mathbf{y} - \boldsymbol{\Theta}\boldsymbol{\mu}) + (y - \boldsymbol{\Theta}\boldsymbol{\mu})^T\mathbf{L}^{-1}\boldsymbol{\Theta}\cdot(-d\boldsymbol{\mu})$$

$$= (d\boldsymbol{\mu})^T(\beta\mathbf{E}^T\mathbf{E} + \boldsymbol{\Theta}^T\mathbf{L}^{-1}\boldsymbol{\Theta})\boldsymbol{\mu} + \boldsymbol{\mu}^T(\beta\mathbf{E}^T\mathbf{E} + \boldsymbol{\Theta}^T\mathbf{L}^{-1}\boldsymbol{\Theta})d\boldsymbol{\mu} - \beta(d\boldsymbol{\mu})^T\mathbf{E}^T\mathbf{r} - \beta\mathbf{r}^T\,\mathbf{E}\cdot d\boldsymbol{\mu} - (d\boldsymbol{\mu})^T\boldsymbol{\Theta}^T\mathbf{L}^{-1}\mathbf{y} - \mathbf{y}^T\mathbf{L}^{-1}\boldsymbol{\Theta}\cdot d\boldsymbol{\mu}$$

$$= (d\boldsymbol{\mu})^T\boldsymbol{\Sigma}^{-1}\boldsymbol{\mu} + \boldsymbol{\mu}^T\boldsymbol{\Sigma}^{-1}\cdot d\boldsymbol{\mu} - (d\boldsymbol{\mu})^T\boldsymbol{\Sigma}^{-1}\boldsymbol{\mu} - \boldsymbol{\mu}^T\boldsymbol{\Sigma}^{-1}\cdot d\boldsymbol{\mu} = 0 \quad \text{(A13)}$$

## Appendix B: Derivation of conditional posterior PDFs for Algorithm 2

In order to allow the tractability of integration involved later, we incorporate the equation-error precision $\beta$ in the prediction error covariance matrices L by defining $\nu = \beta\kappa$ (which is $\tau, \upsilon$ and $\gamma$ in Subsection 3.2) and $\Phi = \beta L$, so $\Phi = \Phi(\nu)$. We also use similar notation as in Appendix A to give a general form of the results.

The posterior PDF for $\beta$ is readily obtained by using Bayes' theorem:

$$p(\beta|\mathbf{y}, \mathbf{x}_2, \mathbf{x}_3, \boldsymbol{\nu}, b_0) \propto p(\mathbf{y}|\mathbf{x}_2, \mathbf{x}_3, \boldsymbol{\nu}, \beta)p(\mathbf{x}_2, \mathbf{x}_3|\beta)p(\beta|b_0)$$

$$\propto (2\pi/\beta)^{-K/2}\cdot(2\pi/\beta)^{(M-N_dN_m)/2}\cdot|\boldsymbol{\Phi} + \boldsymbol{\Theta}(\mathbf{E}^T\mathbf{E})^{-1}\boldsymbol{\Theta}^T|^{-1/2}\cdot|\mathbf{E}^T\mathbf{E}|^{-1/2}$$

$$\cdot\exp\left\{-\frac{\beta}{2}[(\mathbf{y} - \boldsymbol{\Theta}(\mathbf{E}^T\mathbf{E})^{-1}\mathbf{E}^T\mathbf{r})^T(\boldsymbol{\Phi} + \boldsymbol{\Theta}(\mathbf{E}^T\mathbf{E})^{-1}\boldsymbol{\Theta}^T)^{-1}(\mathbf{y} - \boldsymbol{\Theta}(\mathbf{E}^T\mathbf{E})^{-1}\mathbf{E}^T\mathbf{r}) + (\mathbf{r}^T\mathbf{r} - \mathbf{r}^T\mathbf{E}\,(\mathbf{E}^T\mathbf{E})^{-1}\mathbf{E}^T\mathbf{r})] - b_0\beta\right\}$$

$$\propto (1/\beta)^{(M-N_dN_m-K)/2}\cdot|\boldsymbol{\Phi} + \boldsymbol{\Theta}(\mathbf{E}^T\mathbf{E})^{-1}\boldsymbol{\Theta}^T|^{-\frac{1}{2}}\cdot|\mathbf{E}^T\mathbf{E}|^{-1/2}$$

$$\cdot\exp\left\{-\frac{\beta}{2}[(\mathbf{y} - \boldsymbol{\Theta}\boldsymbol{\mu})^T\boldsymbol{\Phi}^{-1}(\mathbf{y} - \boldsymbol{\Theta}\boldsymbol{\mu}) + ||\mathbf{E}\boldsymbol{\mu} - \mathbf{r}||^2 + 2b_0]\right\}$$

$$\propto \mathrm{Gamma}\left(\beta|a_0^{(1)}, b_0^{(1)}\right). \quad \text{(B1)}$$

where:

$$(\mathbf{y}-\mathbf{\Theta}(\mathbf{E}^T\mathbf{E})^{-1}\mathbf{E}^T\mathbf{r})^T(\mathbf{\Phi}+\mathbf{\Theta}(\mathbf{E}^T\mathbf{E})^{-1}\mathbf{\Theta}^T)^{-1}(\mathbf{y}-\mathbf{\Theta}(\mathbf{E}^T\mathbf{E})^{-1}\mathbf{E}^T\mathbf{r})+(\mathbf{r}^T\mathbf{r}-\mathbf{r}^T\mathbf{E}\,(\mathbf{E}^T\mathbf{E})^{-1}\mathbf{E}^T\mathbf{r})$$

$$=(\mathbf{y}-\mathbf{\Theta\mu})^T\mathbf{\Phi}^{-1}(\mathbf{y}-\mathbf{\Theta\mu})+(\boldsymbol{\mu}-(\mathbf{E}^T\mathbf{E})^{-1}\mathbf{E}^T\mathbf{r})^T\mathbf{E}^T\mathbf{E}(\boldsymbol{\mu}-(\mathbf{E}^T\mathbf{E})^{-1}\mathbf{E}^T\mathbf{r})+(\mathbf{r}^T\mathbf{r}-\mathbf{r}^T\mathbf{E}\,(\mathbf{E}^T\mathbf{E})^{-1}\mathbf{E}^T\mathbf{r})$$

$$=(\mathbf{y}-\mathbf{\Theta\mu})^T\mathbf{\Phi}^{-1}(\mathbf{y}-\mathbf{\Theta\mu})+\|\mathbf{E}\boldsymbol{\mu}-\mathbf{r}\|^2 \quad \text{(B2)}$$

and the shape parameter and rate parameter of the posterior Gamma distribution for $\beta$ are:

$$a_0^{(1)}=(N_dN_m+K-M)/2+1, \quad \text{(B3a)}$$

$$b_0^{(1)}=(\mathbf{y}-\mathbf{\Theta\mu})^T\mathbf{\Phi}^{-1}(\mathbf{y}-\mathbf{\Theta\mu})/2+\|\mathbf{E}\boldsymbol{\mu}-\mathbf{r}\|^2/2+b_0 \quad \text{(B3b)}$$

where $K$ is the dimensionality of data $\mathbf{y}$.

We use the Total Probability Theorem and (A5), (B1) to get the posterior:

$$p(\mathbf{x}_1|\mathbf{y},\mathbf{x}_2,\mathbf{x}_3,\mathbf{v},,b_0)=\int p(\mathbf{x}_1|\mathbf{y},\mathbf{x}_2,\mathbf{x}_3,\mathbf{v},\beta)\,p(\beta|\mathbf{y},\mathbf{x}_2,\mathbf{x}_3,\mathbf{v},,b_0)d\beta$$

$$=\int\mathcal{N}(\mathbf{x}_1|\boldsymbol{\mu},(\beta\mathbf{\Lambda})^{-1})\,\mathrm{Gamma}\left(\beta|a_0^{(1)},b_0^{(1)}\right)d\beta$$

$$=\frac{\Gamma\left(a_0^{(1)}+M/2\right)\left(a_0^{(1)}/b_0^{(1)}\right)^{M/2}}{\Gamma\left(a_0^{(1)}\right)\left(2\pi a_0^{(1)}\right)^{M/2}|\mathbf{\Lambda}|^{-1/2}}\left\{1+\frac{1}{2a_0^{(1)}}\left[\frac{a_0^{(1)}}{b_0^{(1)}}(\mathbf{x}_1-\boldsymbol{\mu})^T\mathbf{\Lambda}\,(\mathbf{x}_1-\boldsymbol{\mu})\right]\right\}^{-M/2-a_0^{(1)}}$$

$$=\mathrm{St}\left(\mathbf{x}_1|\boldsymbol{\mu},\frac{a_0^{(1)}}{b_0^{(1)}}\mathbf{\Lambda},2a_0^{(1)}\right) \quad \text{(B4)}$$

where $\boldsymbol{\mu}=\mathbf{\Lambda}^{-1}(\mathbf{\Theta}^T\mathbf{\Phi}^{-1}\mathbf{y}+\mathbf{E}^T\mathbf{r}),\ \mathbf{\Lambda}=\mathbf{E}^T\mathbf{E}+\mathbf{\Theta}^T\mathbf{\Phi}^{-1}\mathbf{\Theta}$. (B5)

To obtain the MAP values of the hyper-parameters $[\mathbf{v},b_0]$, we need to maximize the following posterior PDF: $p(\mathbf{v},b_0|\mathbf{y},\mathbf{x}_2,\mathbf{x}_3)\propto p(\mathbf{y},\mathbf{x}_2,\mathbf{x}_3|\mathbf{v},b_0)p(\mathbf{v})p(b_0)$. (B6)

We first derive the likelihood function:

$$p(\mathbf{y},\mathbf{x}_2,\mathbf{x}_3|\mathbf{v},b_0)=\int p(\mathbf{y}|\mathbf{x}_2,\mathbf{x}_3,\mathbf{v},\beta)p(\mathbf{x}_2,\mathbf{x}_3|\beta)p(\beta|b_0)d\beta$$

$$=p(\mathbf{y}|\mathbf{x}_2,\mathbf{x}_3,\mathbf{v},\beta)p(\mathbf{x}_2,\mathbf{x}_3|\beta)p(\beta|b_0)/p(\beta|\mathbf{y},\mathbf{x}_2,\mathbf{x}_3,\mathbf{v},b_0)$$

$$=cb_0(2\pi)^{(M-N_dN_m-K)/2}\Gamma((N_dN_m+K-M)/2+1)^{-1}\cdot|\mathbf{\Phi}+\mathbf{\Theta}(\mathbf{E}^T\mathbf{E})^{-1}\mathbf{\Theta}^T|^{-1/2}|\mathbf{E}^T\mathbf{E}|^{-1/2}$$

$$\cdot((\mathbf{y}-\mathbf{\Theta\mu})^T\mathbf{\Phi}^{-1}(\mathbf{y}-\mathbf{\Theta\mu})/2+\|\mathbf{E}\boldsymbol{\mu}-\mathbf{r}\|^2/2+b_0)^{(-(N_dN_m+K-M)/2-1)} \quad \text{(B7)}$$

By assigning broad uniform priors on $\mathbf{v}$ and $b_0$, the objective function for optimizing $\mathbf{v}$ and $b_0$ to find their MAP values $\tilde{\mathbf{v}}$ and $\tilde{b}_0$ becomes the logarithm function of $p(\mathbf{y}|\mathbf{x}_2,\mathbf{x}_3,\mathbf{v},b_0)$ excluding the constants that do not depend on $\mathbf{v}$ and $b_0$:

$$\mathcal{L}(\mathbf{v},b_0)=\log b_0-\frac{1}{2}\log|\mathbf{\Phi}+\mathbf{\Theta}(\mathbf{E}^T\mathbf{E})^{-1}\mathbf{\Theta}^T|$$

$$-((N_dN_m+K-M)/2+1)\log((\mathbf{y}-\mathbf{\Theta\mu})^T\mathbf{\Phi}^{-1}(\mathbf{y}-\mathbf{\Theta\mu})/2+\|\mathbf{E}\boldsymbol{\mu}-\mathbf{r}\|^2/2+b_0) \quad \text{(B8)}$$

The derivative of $\mathcal{L}(\mathbf{v}, b_0)$ with respect to $b_0$ is:

$$\frac{\partial \mathcal{L}(\mathbf{v}, b_0)}{\partial b_0} = \frac{1}{b_0} - \frac{(N_d N_m + K - M + 2)}{(\mathbf{y} - \boldsymbol{\Theta}\boldsymbol{\mu})^T \boldsymbol{\Phi}^{-1} (\mathbf{y} - \boldsymbol{\Theta}\boldsymbol{\mu}) + \|\mathbf{E}\boldsymbol{\mu} - \mathbf{r}\|^2 + 2b_0} \tag{B9}$$

By setting the derivative in (B9) to zero, we get:

$$\tilde{b}_0 = \frac{1}{N_d N_m + K - M} \left( (\mathbf{y} - \boldsymbol{\Theta}\boldsymbol{\mu})^T \boldsymbol{\Phi}^{-1} (\mathbf{y} - \boldsymbol{\Theta}\boldsymbol{\mu}) + \|\mathbf{E}\boldsymbol{\mu} - \mathbf{r}\|^2 \right) \tag{B10}$$

If the derivative of diagonal matrix $\boldsymbol{\Phi}(\mathbf{v})$ with respect to component $v_m$ in $\mathbf{v}$ is $\frac{\partial \boldsymbol{\Phi}(\mathbf{v})}{\partial m} = \mathbf{Z}_m$, then we have $\boldsymbol{\Phi}^{-1}\mathbf{Z}_m = v_m^{-1}\mathbf{Z}_m$ and $\boldsymbol{\Phi}^{-2}\mathbf{Z}_m = v_m^{-2}\mathbf{Z}_m$. The derivative of $\mathcal{L}(\mathbf{v}, b_0)$ with respect to $v_m$ is given by:

$$\frac{\partial \mathcal{L}(\mathbf{v}, b_0)}{\partial v_m} = -\frac{1}{2} \frac{\partial \left( \log \left| \boldsymbol{\Phi} + \boldsymbol{\Theta} (\mathbf{E}^T \mathbf{E})^{-1} \boldsymbol{\Theta}^T \right| \right)}{\partial v_m} - ((N_d N_m + K - M)/2 + 1) \frac{\frac{\partial}{\partial v_m} [(\mathbf{y} - \boldsymbol{\Theta}\boldsymbol{\mu})^T \boldsymbol{\Phi}^{-1} (\mathbf{y} - \boldsymbol{\Theta}\boldsymbol{\mu})/2]}{(\mathbf{y} - \boldsymbol{\Theta}\boldsymbol{\mu})^T \boldsymbol{\Phi}^{-1} (\mathbf{y} - \boldsymbol{\Theta}\boldsymbol{\mu}) + \|\mathbf{E}\boldsymbol{\mu} - \mathbf{r}\|^2 + 2b_0}$$

$$= -\frac{1}{2} \left[ \frac{\partial \log |\boldsymbol{\Lambda}|}{\partial v_m} + \frac{\partial \log |\boldsymbol{\Phi}|}{\partial v_m} + \frac{(K + N_d N_m - M + 2)(\mathbf{y} - \boldsymbol{\Theta}\boldsymbol{\mu})^T \frac{\partial \boldsymbol{\Phi}^{-1}}{\partial v_m} (\mathbf{y} - \boldsymbol{\Theta}\boldsymbol{\mu})}{(\mathbf{y} - \boldsymbol{\Theta}\boldsymbol{\mu})^T \boldsymbol{\Phi}^{-1} (\mathbf{y} - \boldsymbol{\Theta}\boldsymbol{\mu}) + \|\mathbf{E}\boldsymbol{\mu} - \mathbf{r}\|^2 + 2b_0} \right]$$

$$= -\frac{1}{2} \left[ -\mathrm{tr}(\boldsymbol{\Lambda}^{-1} \boldsymbol{\Theta}^T \boldsymbol{\Phi}^{-2} \mathbf{Z}_m \boldsymbol{\Theta}) + \mathrm{tr}(\boldsymbol{\Phi}^{-1} \mathbf{Z}_m) - \frac{(N_d N_m + K - M + 2)(\mathbf{y} - \boldsymbol{\Theta}\boldsymbol{\mu})^T \boldsymbol{\Phi}^{-2} \mathbf{Z}_m (\mathbf{y} - \boldsymbol{\Theta}\boldsymbol{\mu})}{(\mathbf{y} - \boldsymbol{\Theta}\boldsymbol{\mu})^T \boldsymbol{\Phi}^{-1} (\mathbf{y} - \boldsymbol{\Theta}\boldsymbol{\mu}) + \|\mathbf{E}\boldsymbol{\mu} - \mathbf{r}\|^2 + 2b_0} \right]$$

$$= -\frac{1}{2} \left[ -v_m^{-2} \mathrm{tr}(\boldsymbol{\Lambda}^{-1} \boldsymbol{\Theta}^T \mathbf{Z}_m \boldsymbol{\Theta}) + v_m^{-1} \mathrm{tr}(\mathbf{Z}_i) - \frac{v_m^{-2} (N_d N_m + K - M + 2)(\mathbf{y} - \boldsymbol{\Theta}\boldsymbol{\mu})^T \mathbf{Z}_m (\mathbf{y} - \boldsymbol{\Theta}\boldsymbol{\mu})}{(\mathbf{y} - \boldsymbol{\Theta}\boldsymbol{\mu})^T \boldsymbol{\Phi}^{-1} (\mathbf{y} - \boldsymbol{\Theta}\boldsymbol{\mu}) + \|\mathbf{E}\boldsymbol{\mu} - \mathbf{r}\|^2 + 2b_0} \right] \tag{B11}$$

where: $|\boldsymbol{\Phi} + \boldsymbol{\Theta}(\mathbf{E}^T\mathbf{E})^{-1}\boldsymbol{\Theta}^T| = |\boldsymbol{\Phi}||(\mathbf{E}^T\mathbf{E})^{-1}||\mathbf{E}^T\mathbf{E} + \boldsymbol{\Theta}^T\boldsymbol{\Phi}^{-1}\boldsymbol{\Theta}\ | = |\boldsymbol{\Phi}||(\mathbf{E}^T\mathbf{E})^{-1}||\boldsymbol{\Lambda}|$ (B12)

As in the case of (A12), the terms involving derivatives of $\boldsymbol{\mu}$ drop out when differentiating (B11).
Setting the derivative in (B13) to zero leads to the update formula for the MAP value $\tilde{v}_m$:

$$\tilde{v}_m = \frac{1}{\mathrm{tr}(\mathbf{Z}_m)} \left[ \mathrm{tr}(\boldsymbol{\Lambda}^{-1} \boldsymbol{\Theta}^T \mathbf{Z}_m \boldsymbol{\Theta}) + \frac{(N_d N_m + K - M + 2)(\mathbf{y} - \boldsymbol{\Theta}\boldsymbol{\mu})^T \mathbf{Z}_m (\mathbf{y} - \boldsymbol{\Theta}\boldsymbol{\mu})}{(\mathbf{y} - \boldsymbol{\Theta}\boldsymbol{\mu})^T \boldsymbol{\Phi}^{-1} (\mathbf{y} - \boldsymbol{\Theta}\boldsymbol{\mu}) + \|\mathbf{E}\boldsymbol{\mu} - \mathbf{r}\|^2 + 2b_0} \right]$$

$$= \frac{1}{\mathrm{tr}(\mathbf{Z}_m)} \left[ \mathrm{tr}(\boldsymbol{\Lambda}^{-1} \boldsymbol{\Theta}^T \mathbf{Z}_m \boldsymbol{\Theta}) + \frac{a_0^{(1)} (\mathbf{y} - \boldsymbol{\Theta}\boldsymbol{\mu})^T \mathbf{Z}_m (\mathbf{y} - \boldsymbol{\Theta}\boldsymbol{\mu})}{b_0^{(1)}} \right] \tag{B13}$$

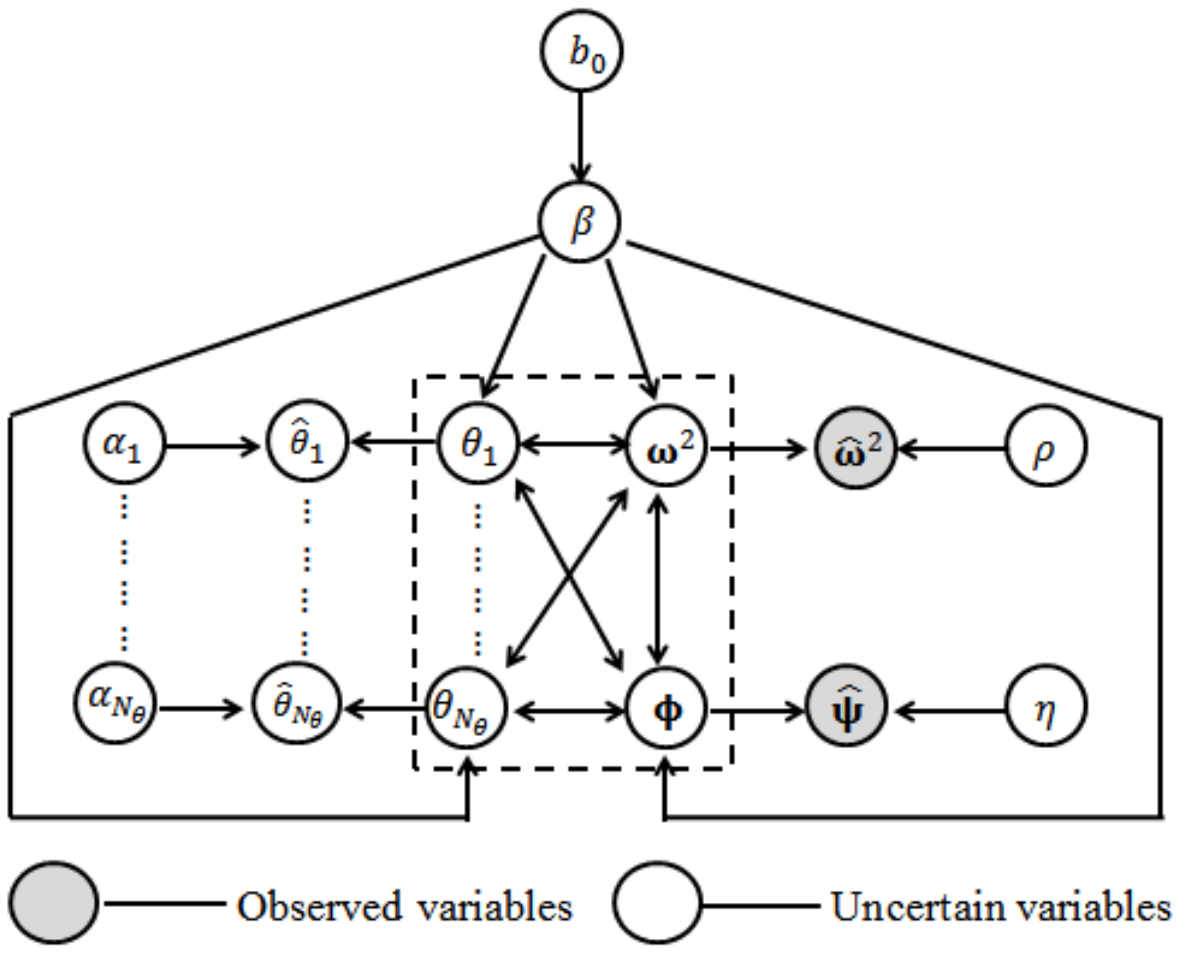

**Fig 1.** Hierarchical Bayesian model representation of the structural system identification problem, where the joint probability model of the model parameters in the dashed box is of interest and each arrow denotes the conditional dependencies used in the hierarchical model.

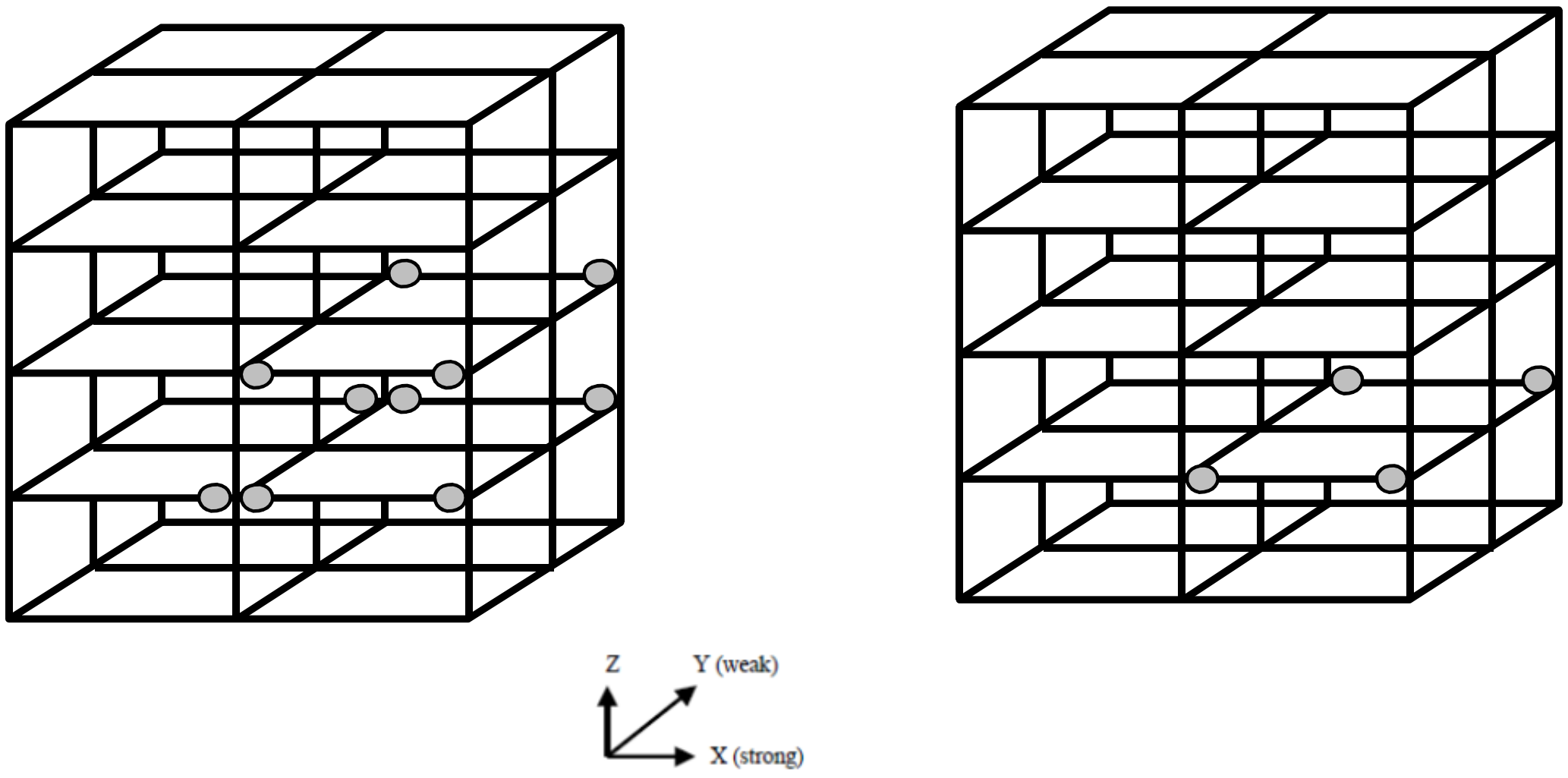

(a) DP1U (b) DP2U

**Fig. 2.** The diagram of the unbraced Phase II benchmark structure, with the damage patterns: (a) DP1U and (b) DP2U. The circles indicate beam-column joint connections with reduced rotational stiffness [30].

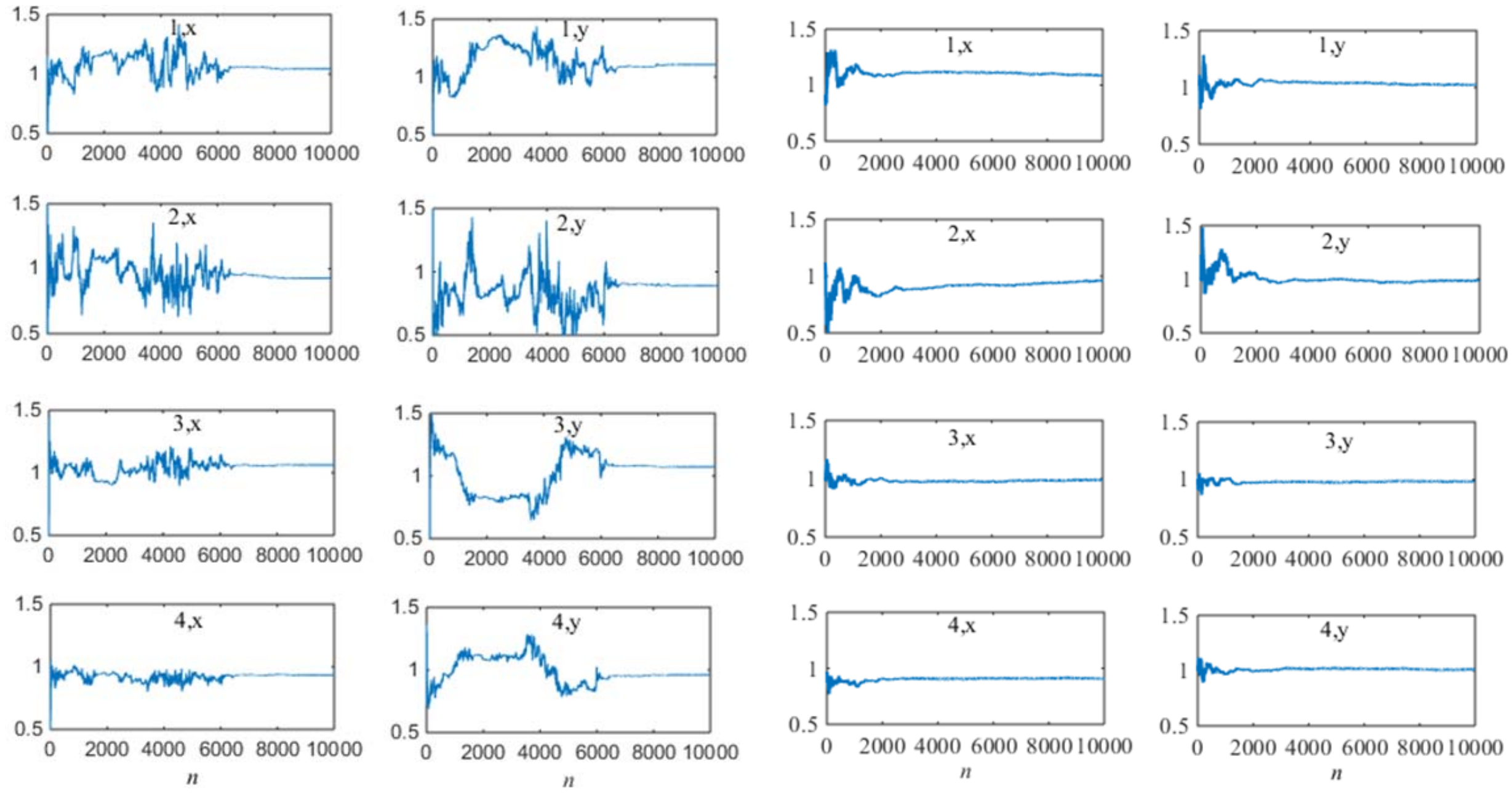

(a) Algorithm 1 (b) Algorithm 2

**Fig. 3.** Unbraced cases in simulated Phase II benchmark: Markov chain samples for the stiffness parameters in the calibration stage, by running: (a) Algorithm 1, and (b) Algorithm 2. The index *n* denotes the time step of the Markov chain.

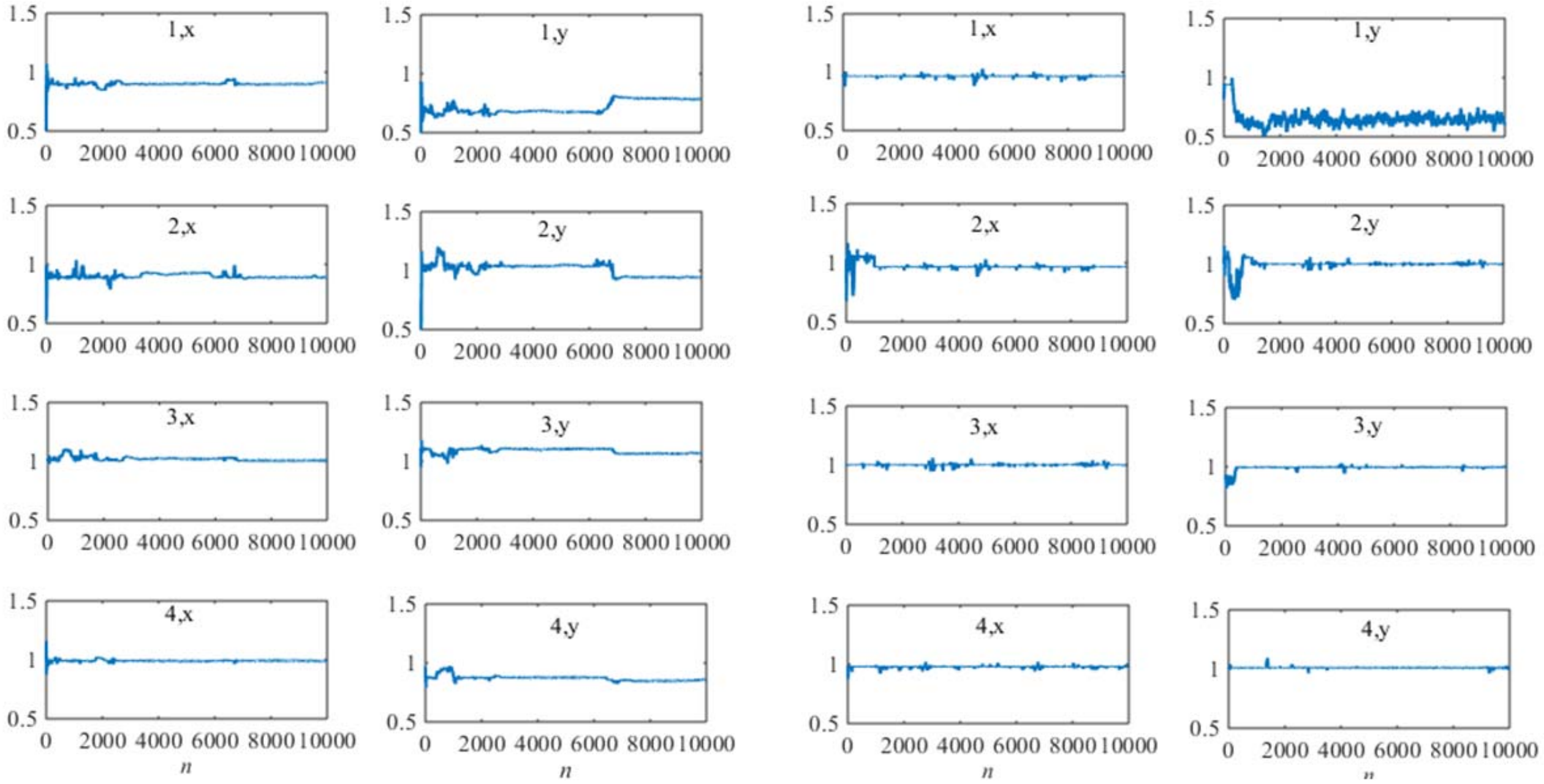

(a) Algorithm 1 (b) Algorithm 2

**Fig. 4.** Unbraced cases in simulated Phase II benchmark: Markov chain samples for the stiffness parameters for the DP2U case, by running: (a) Algorithm 1, and (b) Algorithm 2. The index *n* denotes the time step of the Markov chain.

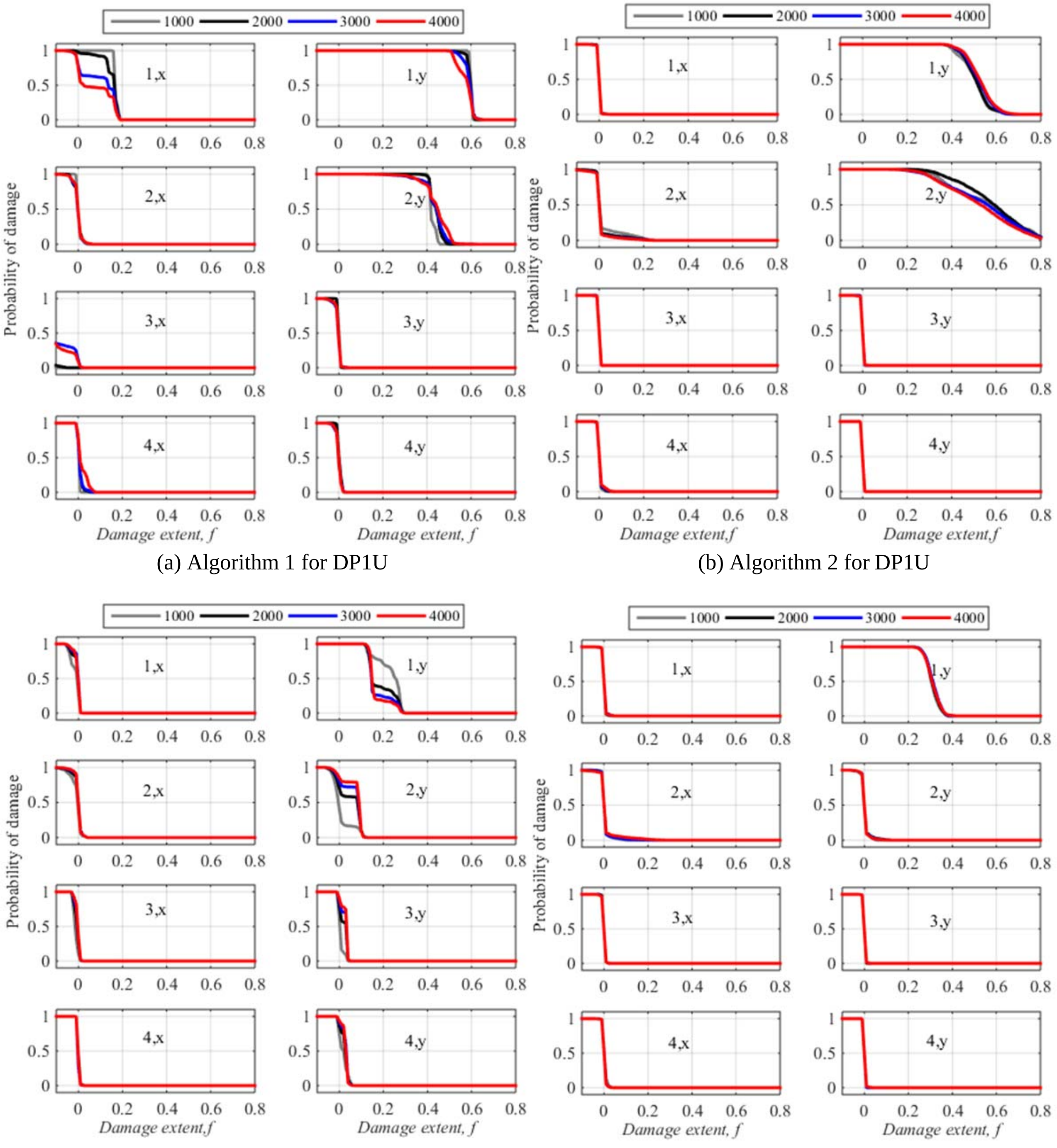

(a) Algorithm 1 for DP1U

(b) Algorithm 2 for DP1U

(c) Algorithm 1 for DP2U

(d) Algorithm 2 for DP2U

**Fig. 5.** Unbraced cases in simulated Phase II benchmark: Estimated damage probability curves for each substructure for damage scenarios: (a)(b) DP1U; (c)(d) DP2U, by running: (a)(c) Algorithm 1; (b)(d) Algorithm 2, using 1,000, 2,000, 3,000, and 4,000 post burn-in samples.

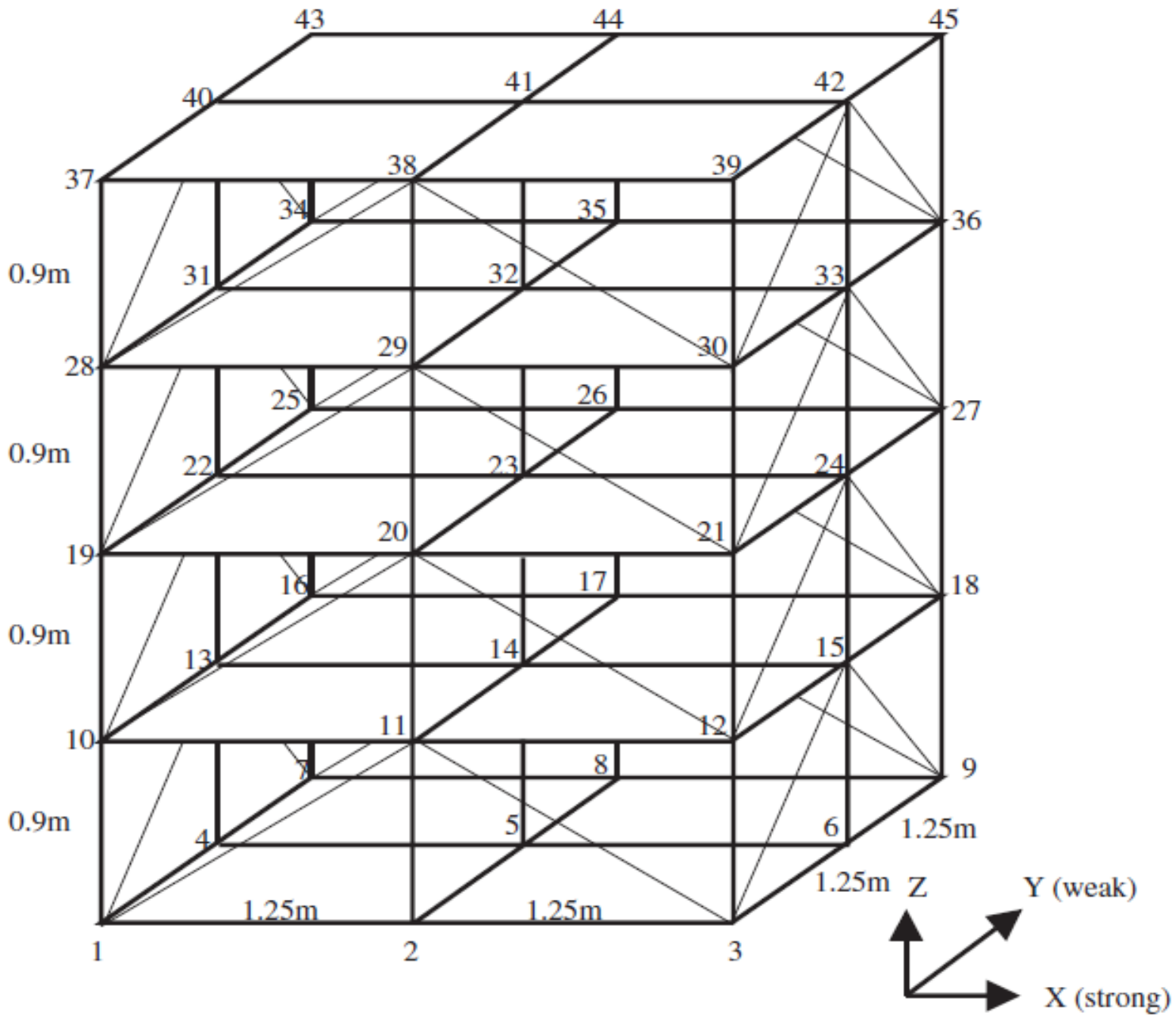

**Fig. 6.** The diagram of the braced benchmark structure [33].

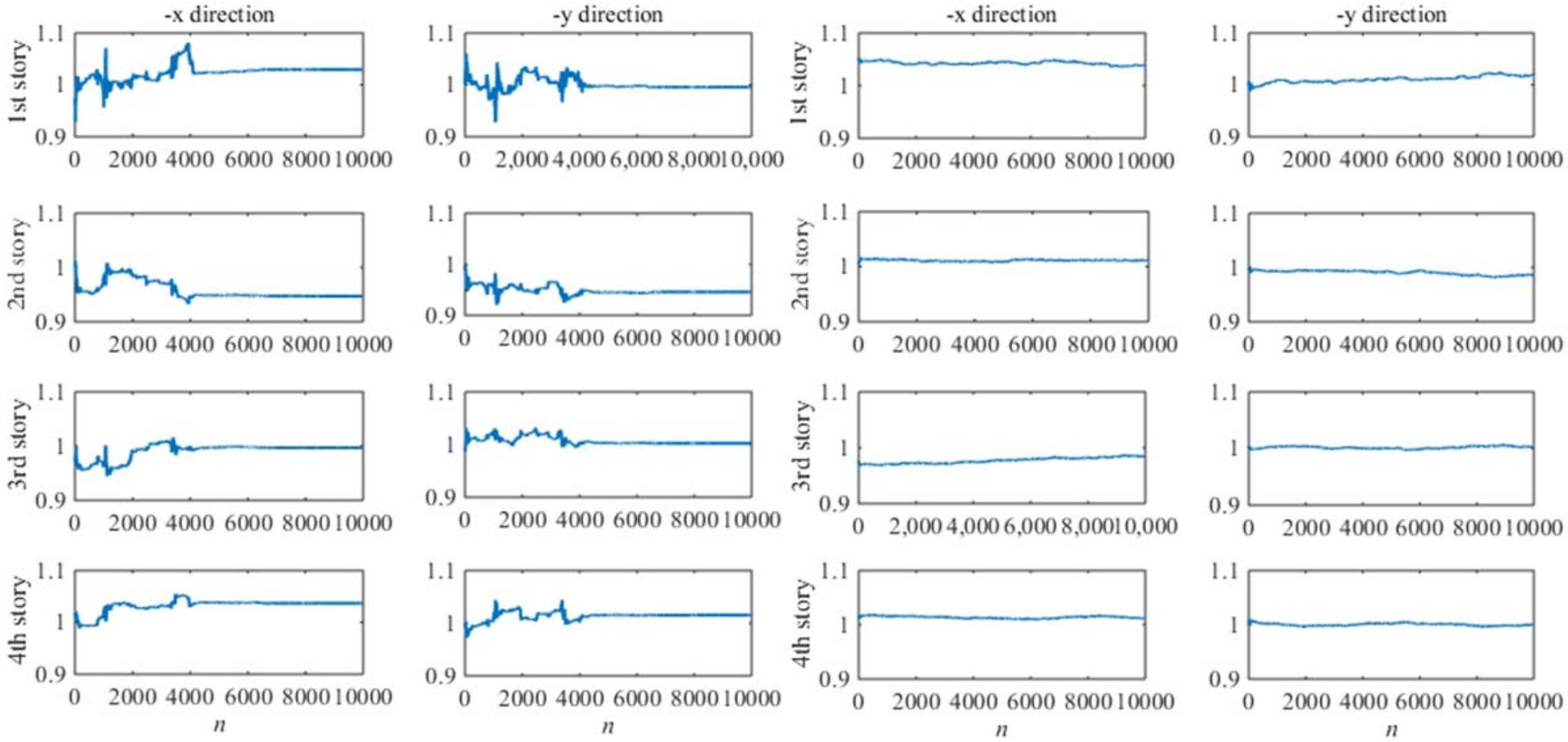

(a) Algorithm 1 (b) Algorithm 2

**Fig. 7.** Brace cases in simulated Phase II benchmark: Posterior Markov chain samples for 8 of the 16 updated stiffness scaling parameters for the RB case (calibration stage), by running: (a) Algorithm 1; (b) Algorithm 2.

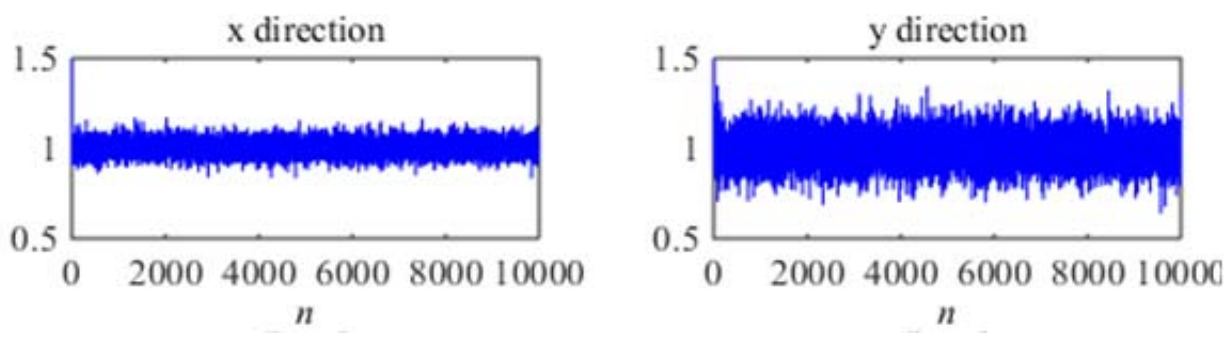

(a) Algorithm 1

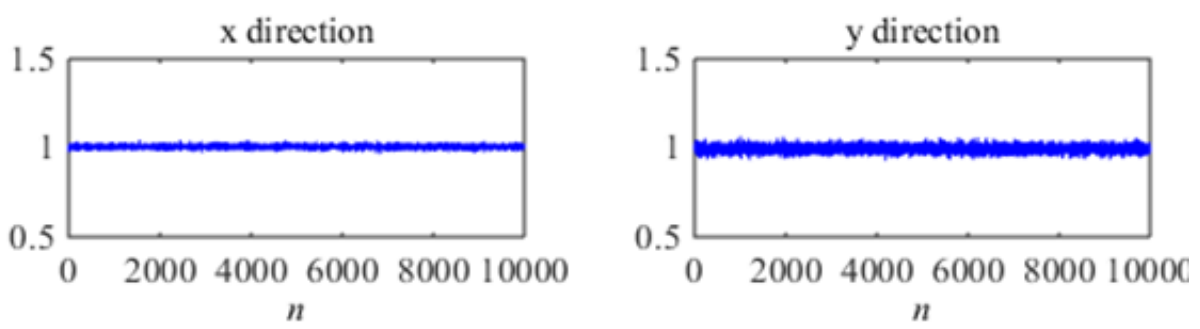

(b) Algorithm 2

**Fig. 8.** Brace cases in experimental Phase II benchmark: Posterior Markov chain samples for the 2 updated stiffness scaling parameters for the Config. 1 case (calibration stage), by running: (a) Algorithm 1; (b) Algorithm 2.

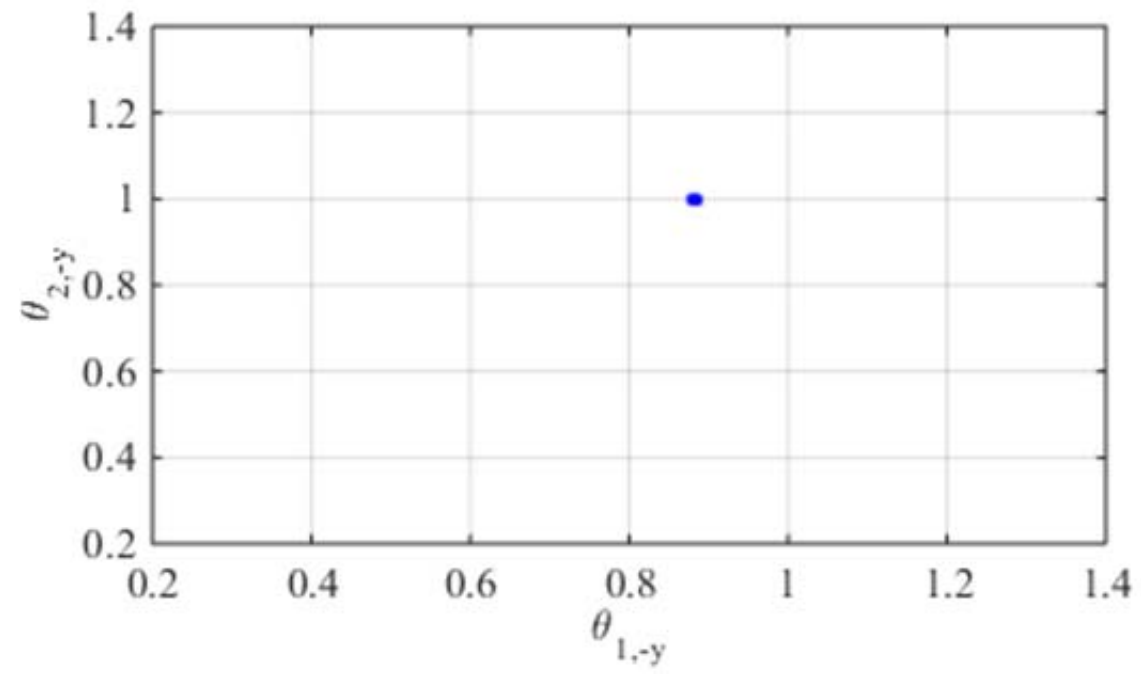

(a) Algorithm 1

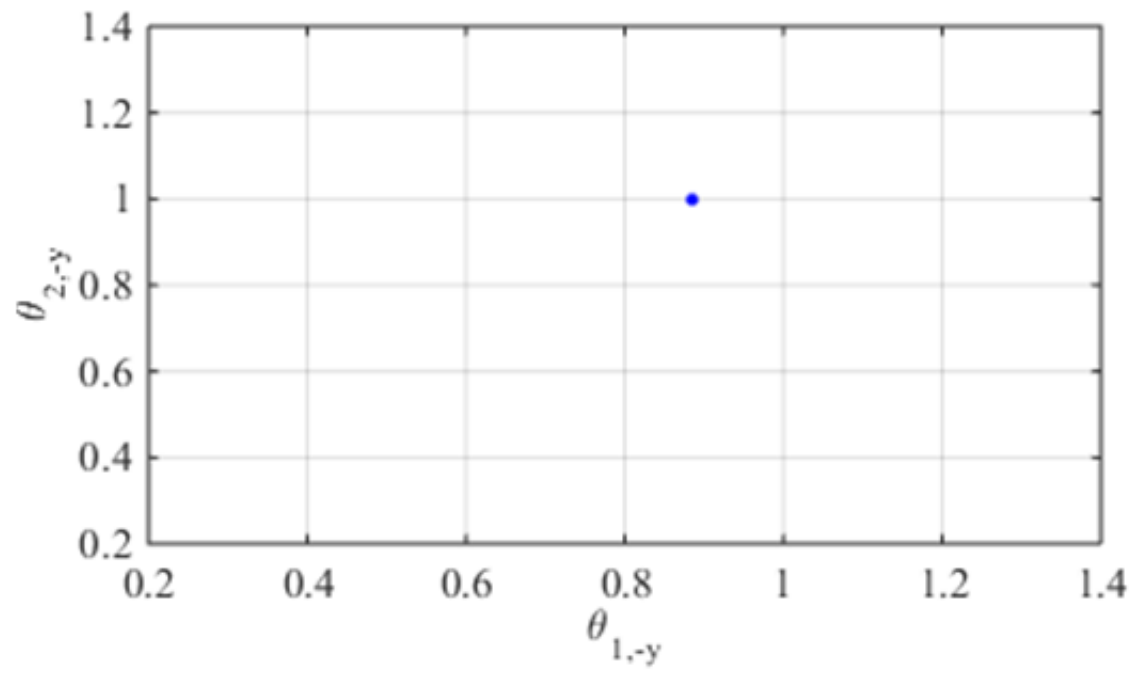

(b) Algorithm 2

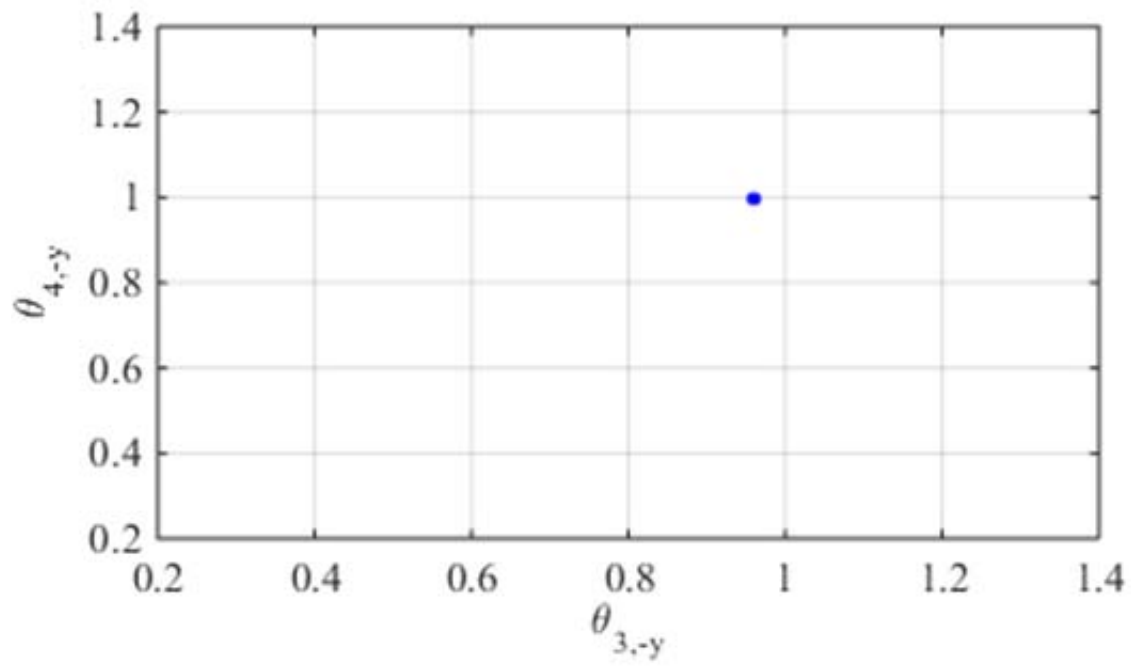

(c) Algorithm 1

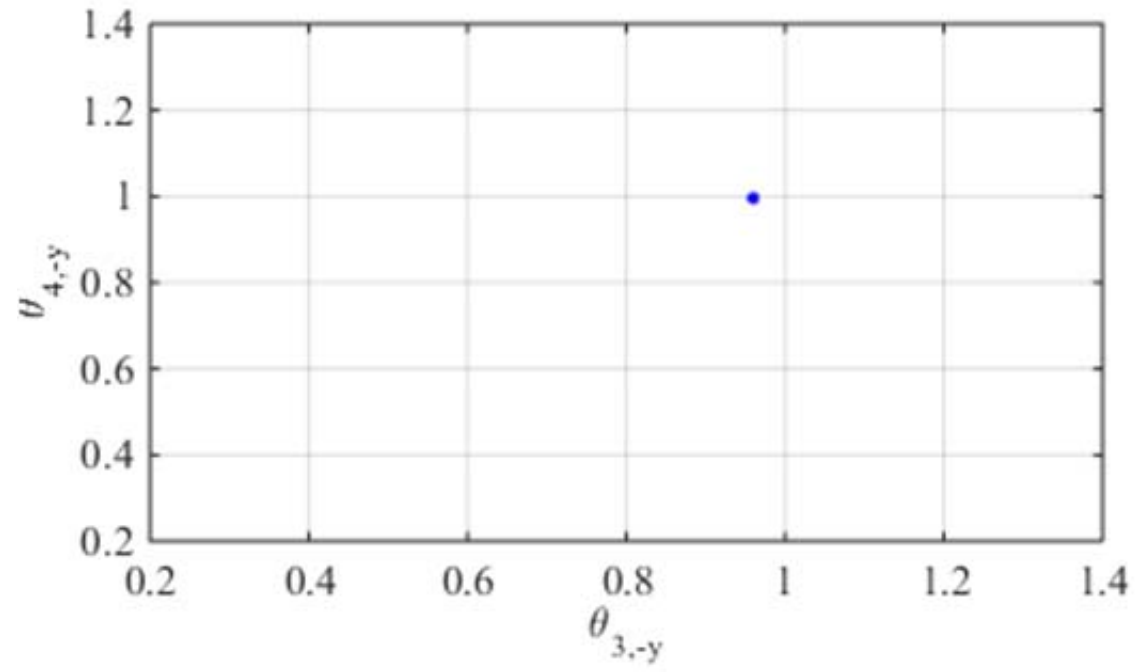

(d) Algorithm 2

**Fig. 9.** Brace cases in simulated Phase II benchmark: Post burn-in samples for posterior stiffness parameter for the DP3B case plotted in: (a)(b) $\{\theta_{1,-y}, \theta_{2,-y}\}$; (c)(d) $\{\theta_{3,-y}, \theta_{4,-y}\}$ spaces, by runing: (a)(c) Algorithm 1; (b)(d) Algorithm 2.

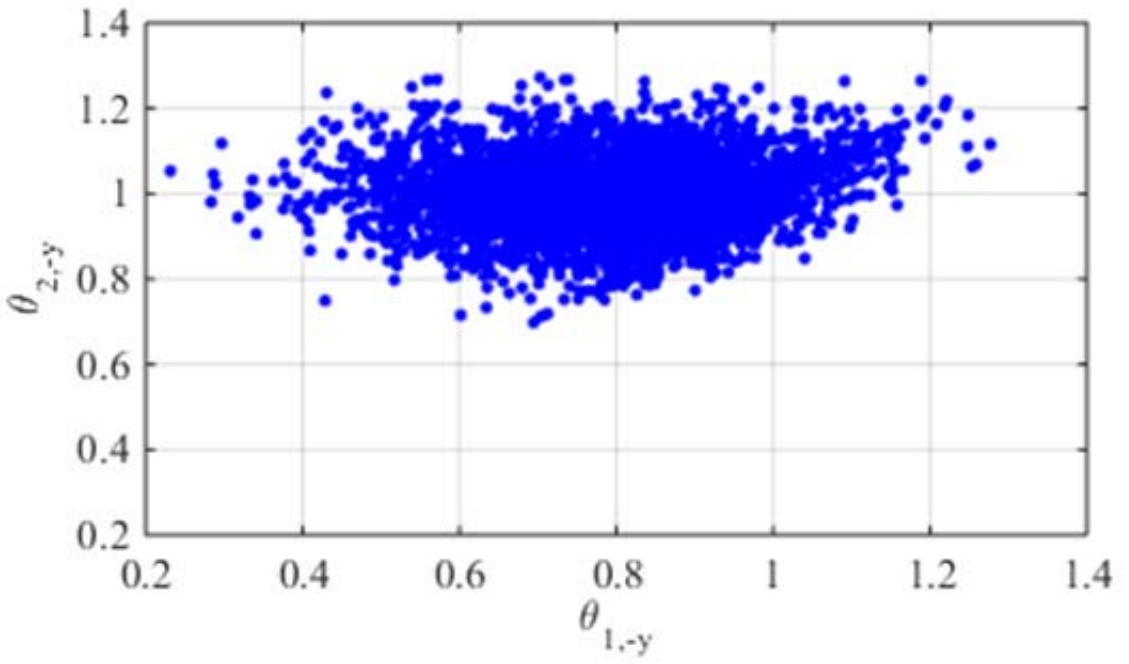

(a) Algorithm 1

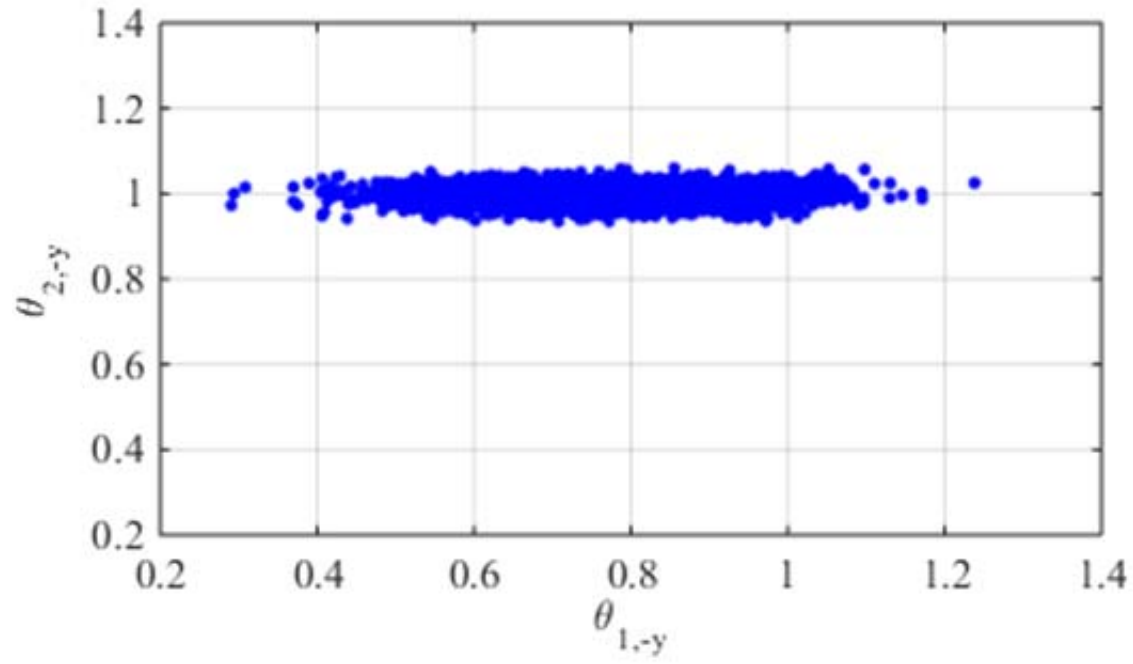

(b) Algorithm 2

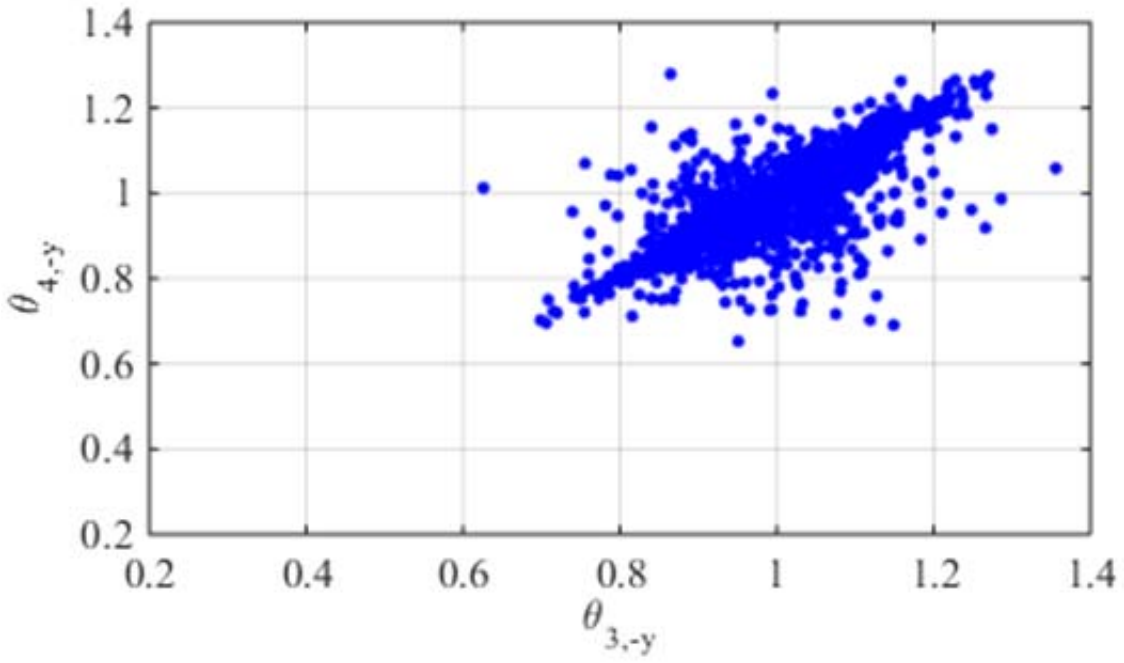

(c) Algorithm 1

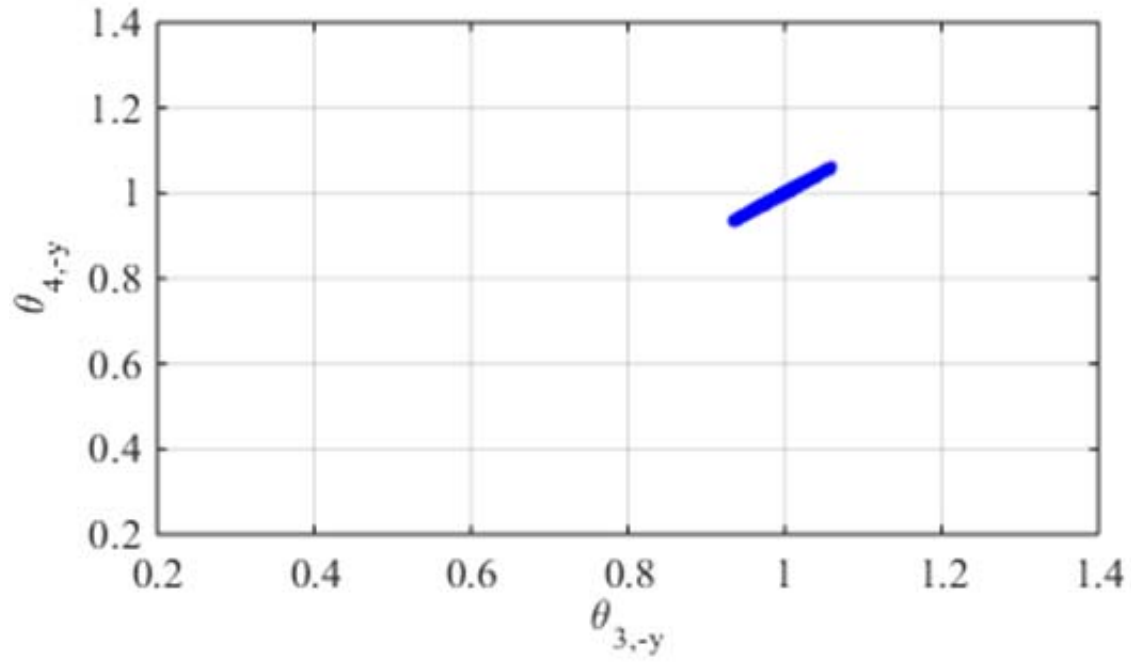

(d) Algorithm 2

**Fig. 10.** Brace cases in experimental Phase II benchmark: Post burn-in samples for posterior stiffness parameter for the Config 5 scenario, plotted in: (a)(b) $\{\theta_{1,-y}, \theta_{2,-y}\}$; (c)(d) $\{\theta_{3,-y}, \theta_{4,-y}\}$ spaces, by running: (a)(c) Algorithm 1; (b)(d) Algorithm 2.

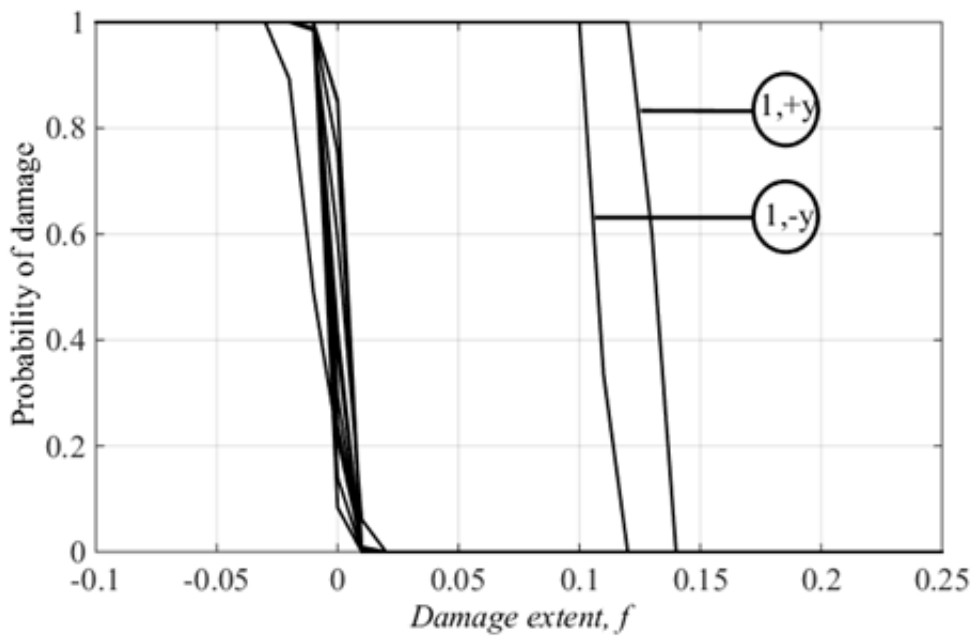

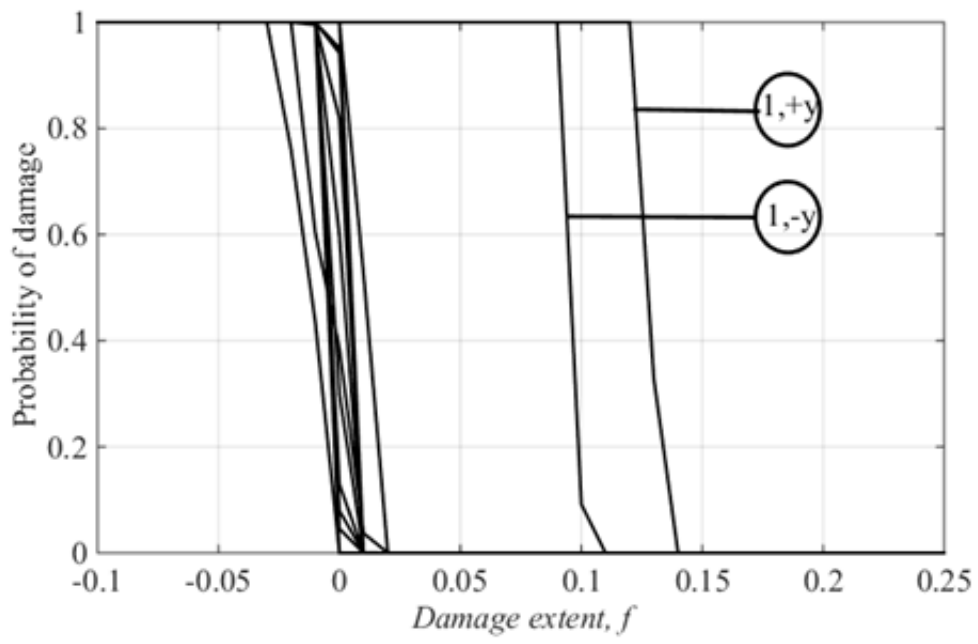

(i) (i)

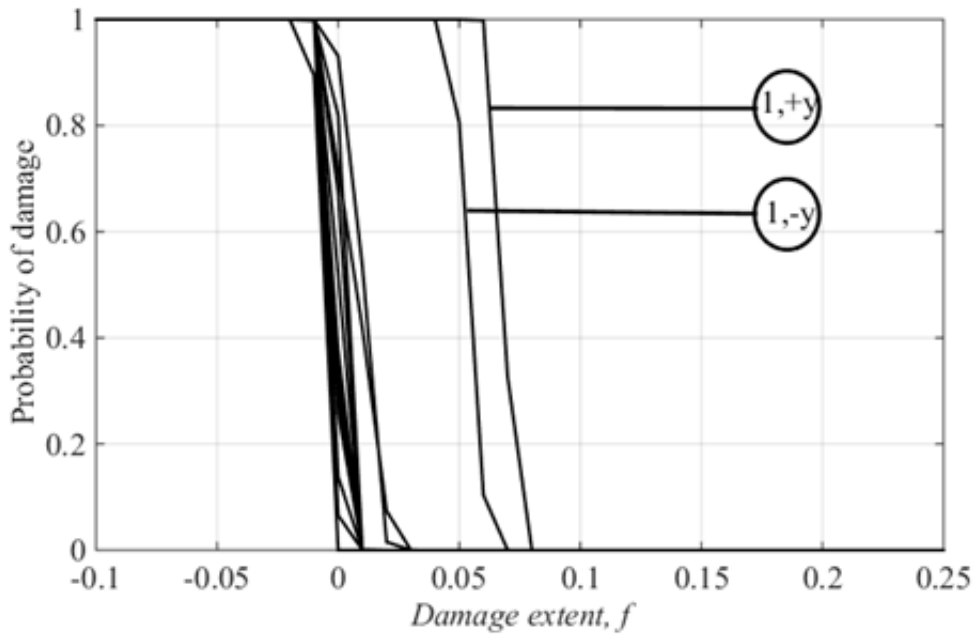

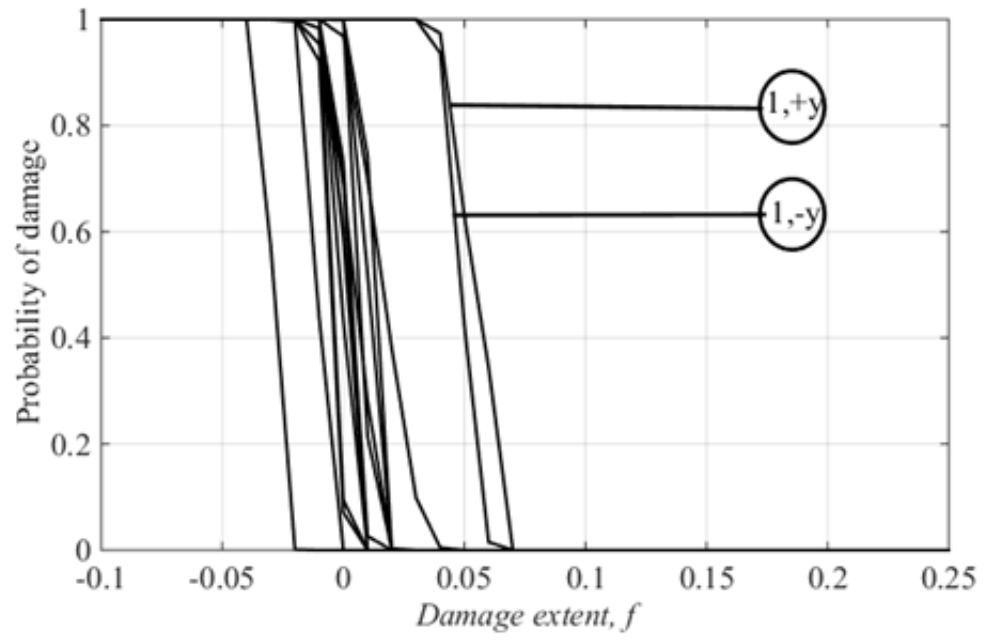

(ii) (ii)

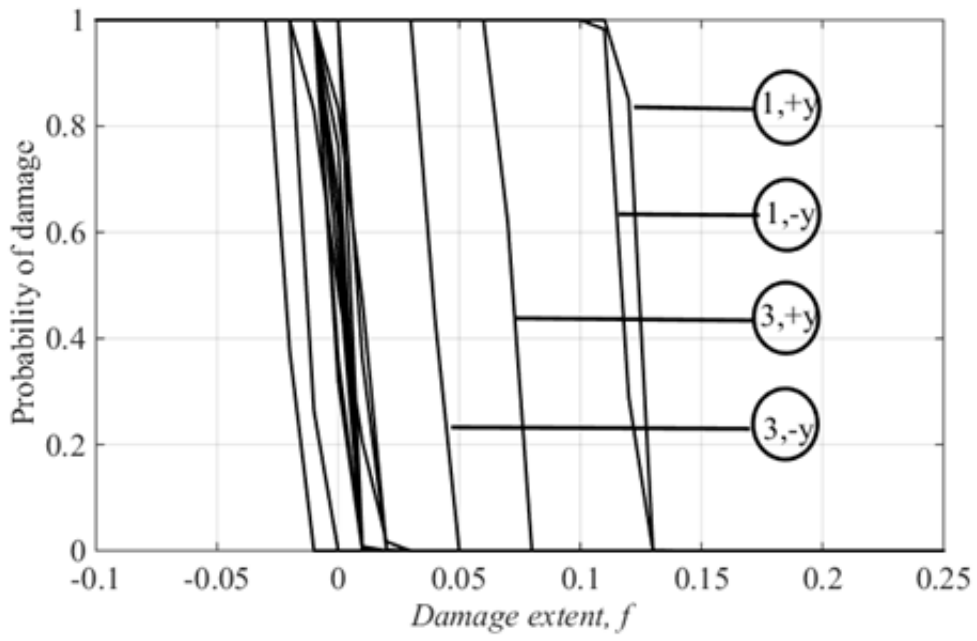

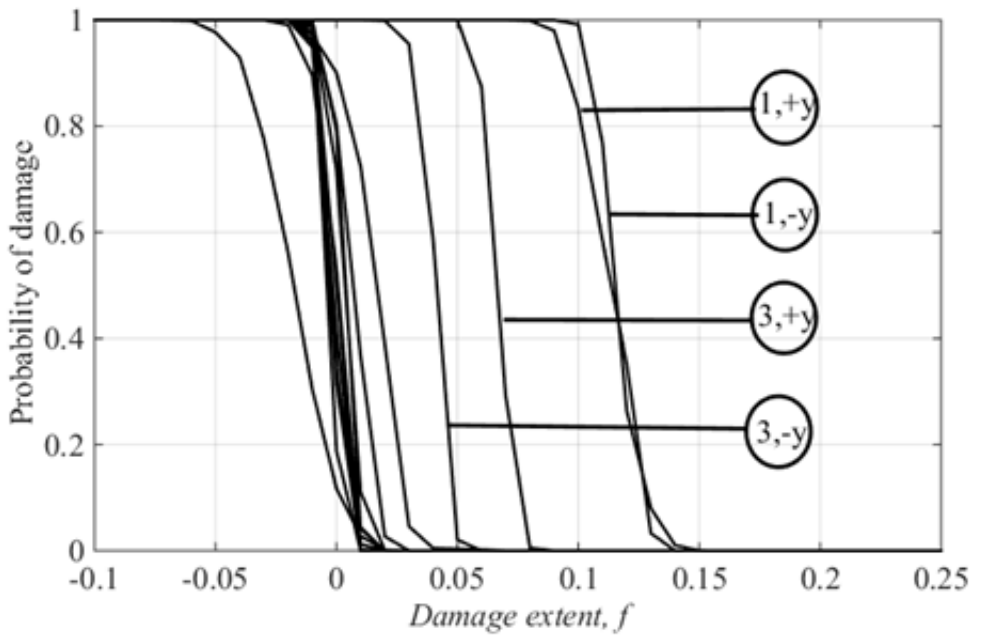

(iii) (iii)

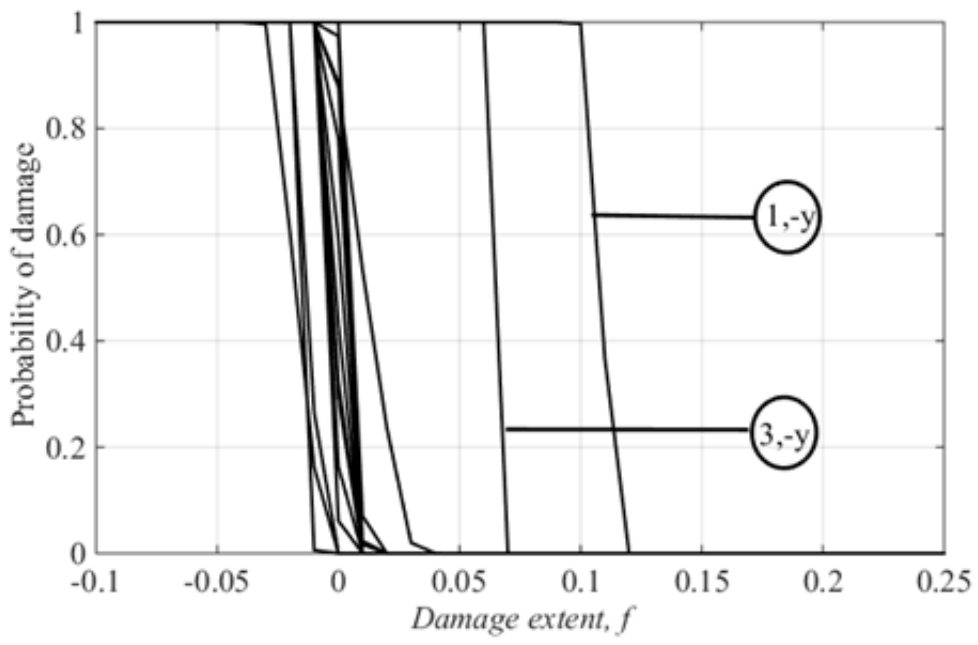

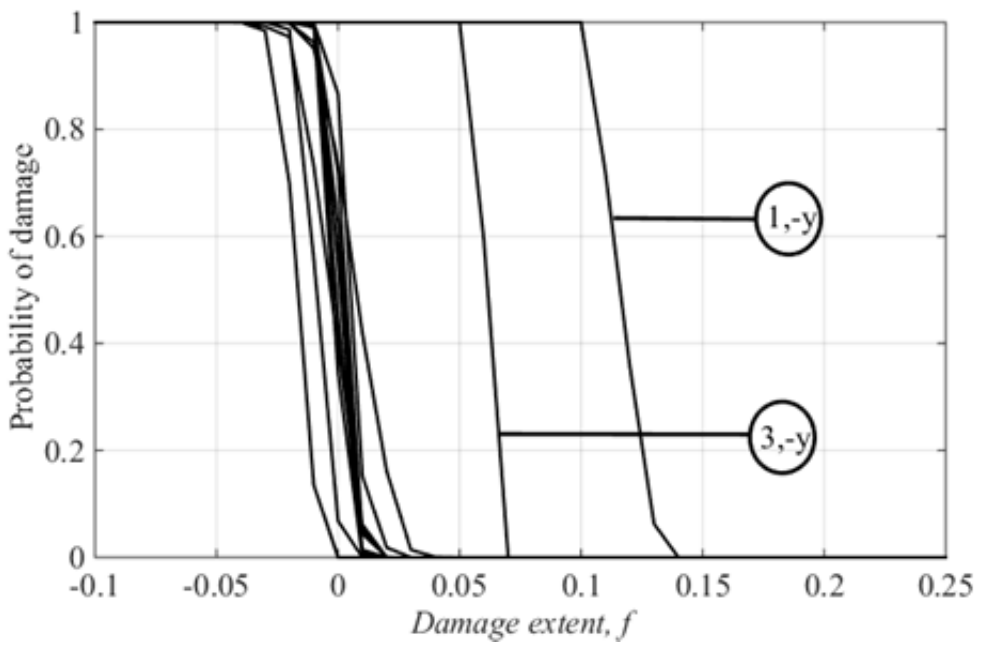

(iv) (iv)

(a) (b)

**Fig. 11***:* Brace cases in simulated Phase II benchmark problem: estimated damage probability curves by (36) for each substructure (one for each of the four faces of each of the four stories) of the scenario: (i) DP1B; (ii) DP2B; (iii) DP3B and (iv) DP3Bu, by running: (a) Algorithm 1; (b) Algorithm 2, using 4,000 post burn-in samples.

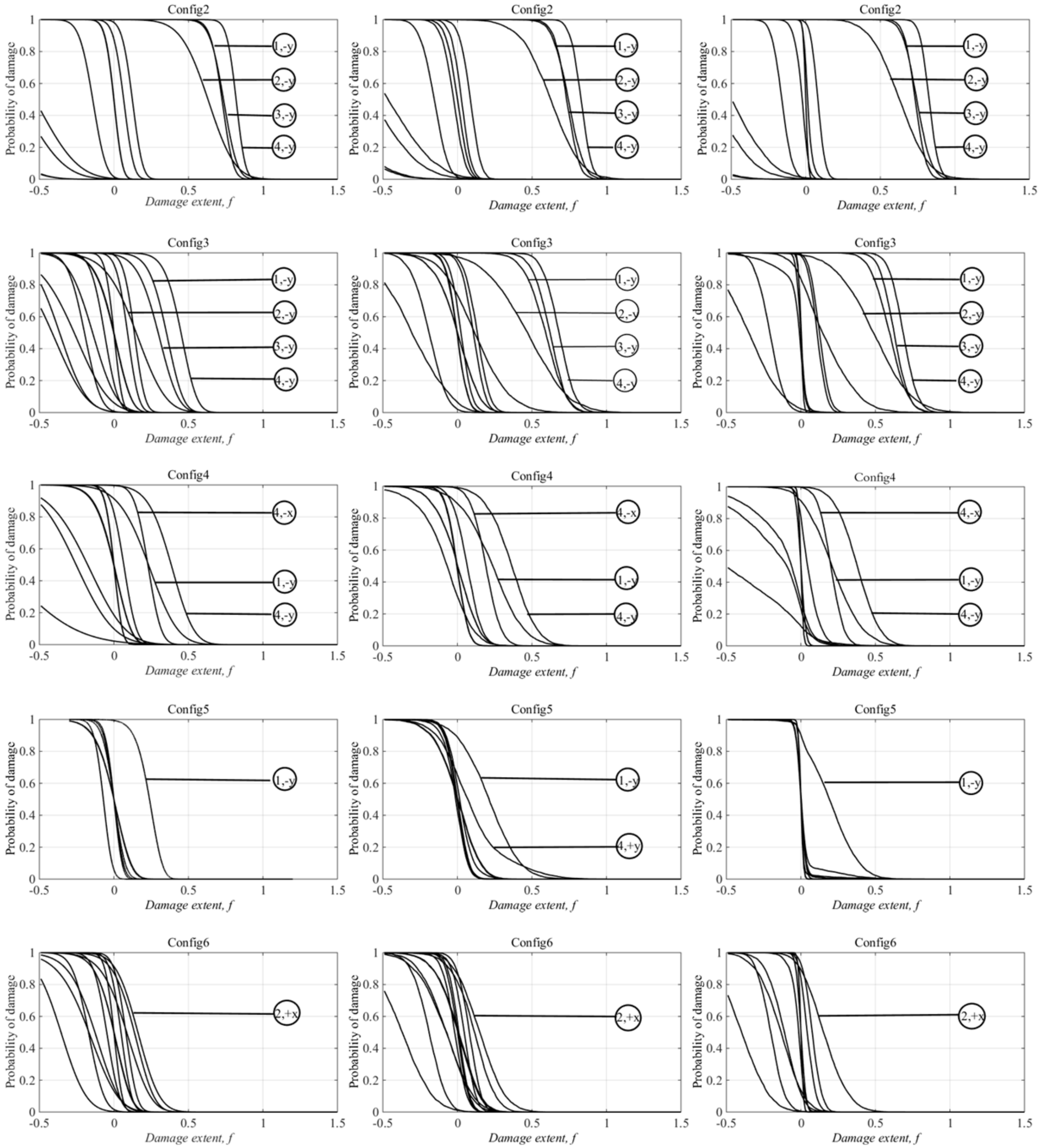

(a) Huang et al. [15] (b) Algorithm 1 (c) Algorithm 2

**Fig. 12.** Brace cases in experimental Phase II benchmark problem: Estimated damage probability curves by (36) for each substructure (one for each of the four faces of each of the four stories) by running: (a) the method in [15]; (b) Algorithm 1; (c) Algorithm 2, using 4,000 post burn-in samples.